\documentclass{cas-dc}
\usepackage[numbers]{natbib}

\usepackage[framemethod=tikz]{mdframed}
\usepackage{graphicx}
\usepackage{subcaption}

\usepackage{amsmath,amssymb,amsfonts}
\usepackage{algorithmic}
\usepackage[ruled, linesnumbered, lined, boxed, commentsnumbered]{algorithm2e} %
\usepackage{color}
\usepackage{tcolorbox} 
\newcommand{\rqbox}[1]{\begin{tcolorbox}[left=4pt,right=4pt,top=4pt,bottom=4pt,colback=gray!5,colframe=gray!40!black,before skip=6pt,after skip=6pt]#1\end{tcolorbox}}

\usepackage{listings}
\usepackage{xcolor}

\SetKwFunction{Fn}{Fn}
\SetKw{Continue}{continue}
\SetKwProg{Fn}{Function}{:}{end}
\newcommand{\MyCall}[2]{#1(#2)}

\def\tsc#1{\csdef{#1}{\textsc{\lowercase{#1}}\xspace}}
\tsc{WGM}
\tsc{QE}
\tsc{EP}
\tsc{PMS}
\tsc{BEC}
\tsc{DE}

\begin{document}
\let\WriteBookmarks\relax
\def\floatpagepagefraction{1}
\def\textpagefraction{.001}
\shorttitle{Journal of Systems and Software}
\shortauthors{M. Zhang, L. Zhou, B. Xiao et~al.}

\title [mode = title]{Understanding and Improving Automated Proof Synthesis for Interactive Theorem Provers}                 

\author[1]{Manqing Zhang}
\fnmark[1]
\fntext[1]{This work was done when the first author was a visiting scholar at Southern University of Science and Technology.}

\author[1]{Yunwei Dong}[orcid=0000-0001-9882-9121]
\cormark[1]

\cortext[1]{Corresponding author}
\ead{yunweidong@nwpu.edu.cn}

\credit{Conceptualization of this study, Methodology, Software}
\address[1]{School of Software, Northwestern Polytechnical University, Xi'an, China}

\author[2]{Lingru Zhou}

\address[2]{ School of Computer Science and Ningbo Institute, Northwestern Polytechnical University, Xi’an, China}

\author[1]{Bingxu Xiao}

\credit{Data curation, Writing - Original draft preparation}

\author[3,4]{Yepang Liu}

\address[3]{Research Institute of Trustworthy Autonomous Systems, Southern University of Science and Technology, Shenzhen, China}
\address[4]{Department of Computer Science and Engineering, Southern University of Science and Technology, Shenzhen, China}

\begin{abstract}
Formal verification using interactive theorem provers ensures high-quality software. However, writing proof scripts for interactive theorem provers is labor-intensive and requires deep expertise. Recent studies have leveraged deep learning to automate theorem proving by learning from manually written proof corpora. Nevertheless, these techniques still achieve limited success rates in proof synthesis. This paper investigates the theorems that current proof synthesis techniques are unable to prove and analyzes their characteristics. Through an in-depth analysis of the proven theorems, proof scripts, and the proof search process, we identify the limitations of existing tools and summarize the key factors influencing proof success rates. Our research provides valuable insights for the future optimization of automated proof tools. Based on our empirical study, we discover that tactic selections conforming to human expert proof patterns are more likely to lead to successful proofs. Inspired by this finding, we propose a pattern-guided tactic search (PGTS) method to heuristically enhance the search process of existing proof synthesis tools. Our evaluation experiments demonstrate that PGTS improves the number of theorems proved by existing proof synthesis tools by an average of 8.05\%, while also achieving an average 20\% increase in previously unproven theorems. Furthermore, PGTS enhances the capability of proof synthesis tools to prove complex theorems and generates more concise proof scripts.
\end{abstract}


\maketitle

\section{Introduction}

Ensuring the reliability of complex software systems remains a fundamental challenge in software engineering~\cite{ferreira2021reliability}, especially given their increasing complexity and critical role in modern applications~\cite{ferreira2023towards,lyu2007software,kapur2011software}. Traditional testing methods often fail to provide complete assurances of software correctness, particularly for edge cases and complex interactions. In this context, interactive theorem provers (ITPs) offers a robust alternative, utilizing formal methods to ensure that software systems meet their specifications. ITPs have been successfully applied to verify many large-scale and complex software systems, such as the seL4 operating system microkernel~\cite{klein2009sel4} and the CompCert compiler~\cite{leroy2009formal}. 

In ITPs, users construct formal proofs by writing proof scripts, consisting of proof tactics sequences. Each tactic represents a small, logical proof step that advances the proof toward completion. However, writing these tactics manually is often a knowledge and labor-intensive task~\cite{staples2014productivity}. For instance, the seL4 team invested over 20 person-years and wrote over 33K lines of proof scripts to verify a microkernel~\cite{zhangsurvey}. Similarly, the CompCert verified compiler project required 6 person-years of effort, resulting in 175 Coq files and approximately 100K lines of proof script~\cite{leroy2016compcert,barthe2019formal}. To alleviate the high cost of manual proof development, recent research has explored automated proof synthesis using neural networks, which learn to generate tactics by training on large-scale corpora of proof step data. These methods treat proof synthesis as a machine translation problem. A model encodes the current proof step state and decodes it into the next appropriate tactic. By iteratively predicting and applying tactics for every proof step state, the system incrementally constructs a complete formal proof through proof search.

\begin{figure}[t]
\centering
\includegraphics[width=\linewidth]{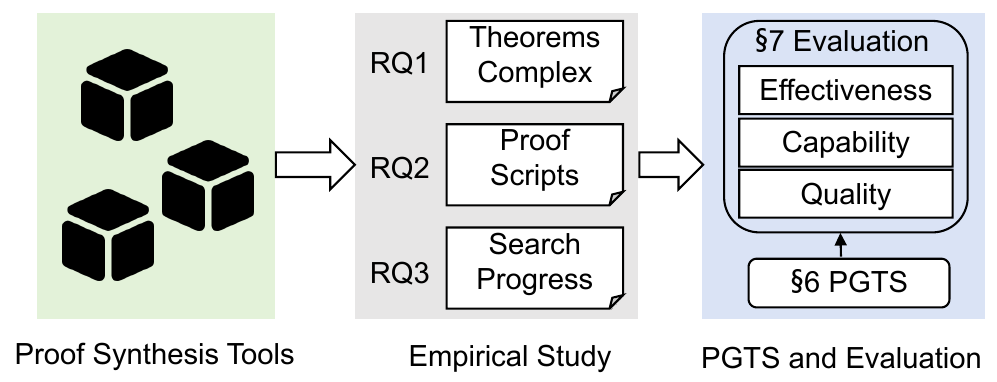}
\caption{Overall Workflow of the Study}
\label{fig:overview}
\end{figure}

However, their performance remains consistently limited in real-world theorem proving applications. To investigate the performance bottlenecks of existing proof synthesis tools, this paper conducts an in-depth analysis of the characteristics of theorems that these tools fail to prove. Specifically, as shown in Figure~\ref{fig:overview}, we examine the features of unprovable theorems from three dimensions: the theorems themselves, the proof scripts, and the proof search process. Our findings indicate that current proof synthesis tools face significant challenges when dealing with complex theorems, especially those involving higher-order logic or intricate logical-mathematical symbols. Proof success rates are notably lower for theorems that include logical operators ($\land$, $\lor$, $\lnot$). Moreover, performance drops substantially when the proof process requires the introduction of intermediate lemmas. We also observe that proof tactics aligned with commonly used human expert reasoning patterns are more likely to succeed in proof search.

Inspired by this study, we propose a pattern-guided tactic search (PGTS) algorithm that heuristically steers the proof search process. As illustrated in Figure~\ref{fig:overview}, our study begins with an empirical investigation into the limitations of existing tools, followed by the design and evaluation of PGTS. This algorithm is a simple yet effective plug-and-play module that integrates seamlessly with existing proof synthesis tools. This idea is also motivated by our observation of existing proof synthesis tool architectures: they typically generate tactics for individual proof steps in isolation, overlooking the correlations and dependencies between tactics across different steps within a proof. Specifically, we utilize a sequential pattern mining algorithm to identify high-frequency proof patterns between proof steps in a proof corpus. These patterns capture common relationships between proof steps that reflect human expert reasoning conventions. By templating these mined patterns, PGTS reranks the candidate tactics produced by proof synthesis tools, prioritizing those that align with the identified patterns. This approach enables the proof search to more closely follow human-like reasoning trajectories. Experimental results on the CoqGym benchmark show that PGTS increases the number of theorems successfully proven by proof synthesis tools by an average of 8.05\%. On average, it proves 20\% more previously unproven theorems. In addition, PGTS is able to handle more complex theorems while generating more concise proofs. All data from the empirical study, the source code and the evaluation results have been released for replication~\cite{pgts}.

This paper makes the following contributions:
\begin{itemize}	
	\item To the best of our knowledge, we conduct the first systematic study on the effectiveness of automated proof synthesis tools. We highlight key limitations of existing techniques and distill the factors influencing proof success rates.
	
	\item Inspired by our empirical findings, we propose PGTS, which leverages human expert proof expertise to heuristically enhance proof search. As a lightweight and plug-and-play approach, PGTS can be seamlessly integrated into the search processes of existing proof synthesis tools, significantly improving their performance.
	
	\item Our experimental evaluation demonstrates that PGTS significantly enhances the performance of proof synthesis tools, achieving an average increase of 8.05\% in successfully proven theorems. Additionally, PGTS improves the capability to prove complex theorems and generates more concise proof scripts, thereby advancing the efficiency and effectiveness of automated proof synthesis.
    
\end{itemize}

The remainder of this paper is organized as follows. Section~\ref{sec:bg} presents the background of this study. Section~\ref{sec:empirical} describes the experimental setup and methodology of our empirical investigation. Section~\ref{sec:study_results} reports the results and findings for each of our research questions. Section~\ref{sec:implications} discusses the implications of these findings in detail. Section~\ref{sec:method} introduces a new method inspired by the findings. Section~\ref{sec:evaluation} presents the experimental evaluation of our proposed method. Section~\ref{sec:discussion} discusses limitations and threats to the validity of our study. Section~\ref{sec:rel} reviews related work. Finally, Section~\ref{sec:con} concludes the paper.
\section{Background}
\label{sec:bg}

This section provides a brief introduction to interactive theorem provers, as well as the workflow of proof synthesis tools.

\subsection{Interactive Theorem Provers}

ITPs are powerful tools widely used in foundational verification tasks, providing a rigorous and structured framework for constructing formal proofs. These systems enable users to define both data and program specifications alongside corresponding proof goals, offering a controlled environment in which proofs can be developed interactively.

In ITPs, the proof process is driven by a sequence of user-issued commands known as tactics. These tactics serve as the fundamental building blocks of interaction, allowing users to structure and guide proofs in a modular and programmable way. Tactics may range from low-level primitives, such as simplification, case analysis, or induction, to higher-level compositions that encapsulate more complex strategies. During proof development, tactics operate by manipulating the current proof context and transforming the original goal into a sequence of simpler subgoals. This process continues iteratively until all subgoals are reduced to axioms or previously proven theorems. While users provide strategic direction by applying tactics, the system is responsible for checking the logical validity of each transformation and offering feedback at every step. In this way, ITPs ensure that all intermediate proof states remain sound under the formal rules of logic.

\begin{figure}[t]
\centering
\includegraphics[width=\linewidth]{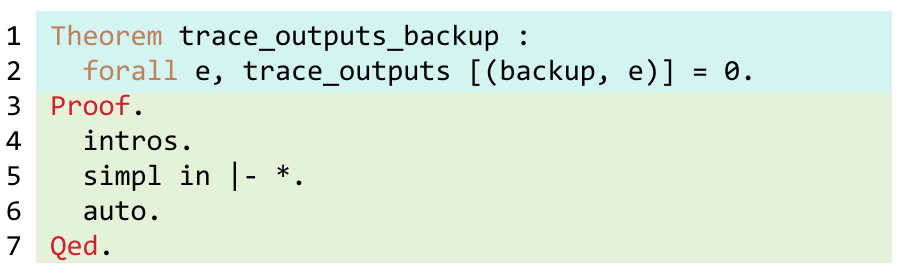}
\caption{An example of proof script in Verdi project}
\label{fig:code}
\end{figure}

Figure~\ref{fig:code} presents an example proof script from the Verdi project, which is concerned with the formal verification of distributed systems. The script defines a theorem named trace\_outputs\_backup, which asserts that any event emitted solely by the backup node does not contribute to the system’s observable output. To establish the correctness of this theorem, the user interacts with the ITPs by applying a sequence of tactics that incrementally construct the proof. As shown in the Figure~\ref{fig:code}, the proof begins with the keyword Proof and concludes with Qed, delineating the start and end of the proof process. The intermediate lines (lines 4-6) represent the specific tactics applied at each step of the derivation. In this case, the script consists of three proof steps, employing the tactics intros, simpl, and auto, respectively. These tactics serve to introduce variables, simplify the goal, and automatically discharge the proof based on the context and available lemmas.



In summary, ITPs support a collaborative, step-by-step construction of formal proofs through the use of tactics. This interactive workflow allows users to tackle complex verification tasks in a modular and verifiable manner, making ITPs indispensable tools for applications that demand high assurance and mathematical rigor.

\subsection{Proof Synthesis Tools}

\begin{figure}[t]
\centering
\includegraphics[width=\linewidth]{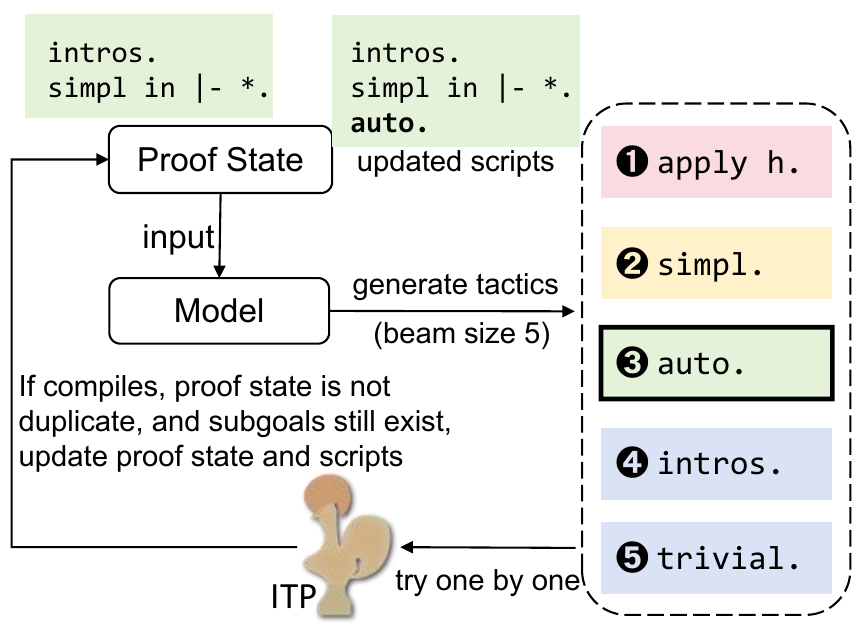}
\caption{An example workflow of step-by-step proof search and tactic selection}
\label{fig:search_example}
\end{figure}

Figure~\ref{fig:search_example} illustrates the workflow of a proof synthesis tool performing proof search in an interactive theorem prover (ITP). To reduce human effort in ITPs, such tools aim to automate the construction of proof scripts for given theorems by mimicking the human proof-writing process. Initially, the proof state contains only the goal, which is the statement of the theorem. The tool encodes this initial state and generates candidate tactics intended to manipulate the state. 

In the Figure~\ref{fig:search_example}, the model predicts five tactics for the current proof state. These tactics are sequentially attempted by the ITP. The first tactic, \texttt{apply h}, fails due to a contextual error. The second, \texttt{simpl}, is syntactically valid but makes no progress, as it reproduces the same proof state. The third, \texttt{auto}, successfully advances the proof by producing an updated proof state. As a result, \texttt{auto} is appended to the proof script, and the updated state, which includes $subgoals$, $local \, context$ of $assumptions$, and $environment$ of proven facts, is passed back to the tool for further prediction. This iterative process continues with the tool generating tactics and the ITP returning updated proof states until there are no more subgoals. At that point, the tool appends \texttt{Qed} to complete the script. If a predefined time or step limit is reached without completing the proof, the search process is terminated.

\section{Empirical Study Methodology}
\label{sec:empirical}

This section describes the experimental setup used in our empirical study, as well as the research questions it addresses, including the motivation and approach.

\subsection{Experimental Setup}

\subsubsection{Selection of Proof Synthesis Tools}

In this study, we selected six state-of-the-art (SOTA) proof synthesis tools, including five based on deep learning and one based on a large language model. Below, we briefly summarize the key characteristics of each tool:

ASTactic~\cite{yang2019learning}: ASTactic employs a standard encoder-decoder architecture that represents proof states as abstract syntax trees (ASTs). It generates syntactically valid proof steps by learning from tree-structured input representations.

Tac~\cite{first2020tactok}: Tac learns to predict proof steps based on the current proof state and the most frequently used Coq tactics from existing partial proof scripts. It excludes custom tactics and ambiguous parameters to reduce learning noise associated with tactic names.

Tok~\cite{first2020tactok}: Tok extends Tac by modeling the proof state together with the full token sequence from prior proof scripts, omitting only punctuation marks. This allows for finer-grained learning of proof construction patterns.

Passport~\cite{sanchez2023passport}: Passport builds upon the Tok framework by additionally encoding identifier information—such as variable and hypothesis names—into the input representation, enhancing the model’s awareness of local proof context.

Proverbot9001~\cite{sanchez2020generating}: Proverbot9001 decomposes the proof step prediction task into two stages: tactic prediction and parameter prediction. The final proof step is synthesized by integrating the outputs of these two components.

PALM~\cite{lu2024proof}: PALM adopts a two-stage synthesis pipeline. First, it leverages ChatGPT to generate an initial complete proof script. This script is then interactively executed in Coq until an error is encountered, at which point a repair mechanism is triggered to correct the script from the failure location onward.

\subsubsection{Dataset}

We use the CoqGym benchmark~\cite{yang2019learning} as the test set for our empirical study. CoqGym is a comprehensive and widely adopted dataset for evaluating proof synthesis tools, containing 13,137 theorems from 27 open-source Coq projects on GitHub. We set up the original CoqGym environment and successfully compiled all 27 projects using Coq 8.9.0. It is worth noting that previous evaluations of Tac, Tok, and Passport were conducted on only 26 projects, as these tools failed to run on the coq-library-undecidability project due to internal Coq errors. In contrast, our experiments successfully executed all tools across the complete set of 27 projects.

Due to version compatibility issues, Proverbot9001 and PALM could not be executed in the original CoqGym environment, as both tools only support newer versions of Coq. To overcome this limitation, we reached out to the authors of these tools to determine whether they had evaluated subsets of the CoqGym benchmark using compatible versions of Coq. The authors of Proverbot9001 provided results obtained on a newer Coq version, covering 11,998 theorems across 27 projects. Similarly, PALM was evaluated on 26 projects within a subset of CoqGym, with results publicly available on GitHub. To ensure a fair comparison across all six tools, we took the intersection of their evaluation sets, resulting in a common subset of 24 projects and 8,879 theorems. This dataset was used to evaluate RQ1 and RQ2, which focus on understanding the limitations of proof synthesis tools from two perspectives: the characteristics of the theorems themselves and human-written proof scripts. For RQ3, which investigates the characteristics of the proof search process itself, such as the number of proof steps, branching behavior, and search depth, we required detailed search trace data. However, such data was not available for Proverbot9001 and PALM. Therefore, we conducted an empirical analysis of the proof search process using only the remaining four tools that provide sufficient dynamic search information.

\subsubsection{Environment}

Evaluating the performance of theorem proving is computationally expensive, but the workload is highly parallelizable, allowing the evaluations to be executed in arbitrary order. For experiments that could be executed on CoqGym, we parallelized the evaluation of all theorems by partitioning them according to project files and running them on a high-performance computing (HPC) server. The HPC environment was managed using the Slurm workload manager. The server was equipped with dual AMD 7285H 32-core processors (totaling 64 cores), 256 GB of DDR4 memory, and a base clock frequency of 2.5 GHz.

\subsection{Research Questions}
\label{sec:state-transition}

\subsubsection{RQ1: Which characteristics of theorems make them harder to prove with existing proof synthesis tools?\\}

\textbf{Motivation:} The structure and characteristics of theorems (such as complexity, logical relationships) play a crucial role in the success rate of proof synthesis tools. Simple theorems may only require basic logical reasoning, whereas more complex theorems often involve challenging features such as recursive structures, higher-order logic, or dependent types. By analyzing the relationship between the characteristics of theorems and the performance of tools, we can identify the limitations of existing proof synthesis tools. This analysis also helps pinpoint their weaknesses when dealing with specific types of theorems. Understanding how these theorem characteristics specifically affect tool performance can provide valuable guidance for optimizing the design of proof synthesis tools.

\textbf{Approach:} To address RQ1, we systematically investigated the characteristics of theorems to understand their impact on proof success rates. Specifically, we analyzed theorem complexity through two key dimensions: (1) the classification of theorems as first-order or higher-order, and (2) the number of logical-mathematical symbols within each theorem.

For the first dimension, we employed CoqHammer to attempt proofs for the given theorems. Theorems successfully proven by CoqHammer were classified as first-order, leveraging CoqHammer’s capability to translate Coq goals into first-order logic. Theorems for which proof attempts failed were categorized as higher-order. This classification provides a robust framework for distinguishing theorem types and analyzing their relationship with proof success rates in automated theorem proving.

For the second dimension, we quantified theorem complexity by counting the occurrences of five types of logical-mathematical symbols: equivalence operators, implication operators, quantifiers, logical operators, and inequality operators. A higher count of these symbols indicated greater theorem complexity. This quantitative approach enabled us to systematically explore how the presence and frequency of logical-mathematical symbols influence proof success rates. By integrating the analysis of theorem type and symbolic complexity with proof outcomes, our study offers nuanced insights into the structural and symbolic factors that shape the challenges of automated theorem proving, highlighting the limitations of current tools and informing strategies for improving their effectiveness.

\subsubsection{RQ2: Which characteristics of proof scripts make them harder to prove with existing proof synthesis tools?\\}

\textbf{Motivation:} The structure and composition of proof scripts play a pivotal role in formal proof construction. A well-structured theorem script enhances readability and maintainability and significantly impacts the proof process's efficiency. Studying the structural characteristics of theorem scripts can reveal critical bottlenecks in the proof process, such as redundant tactic invocations, suboptimal proof strategies, or ineffective lemma decomposition. Identifying these bottlenecks provides valuable insights into how proofs can be streamlined, potentially reducing the time and computational resources required for verification.

\textbf{Approach:} To study RQ2, we summarized the characteristics of proof scripts to investigate their relationship with proof success rates. Specifically, our analysis focused on two key aspects:

1) Lemma Usage: We identified a distinctive feature within proof scripts—the use of lemmas. To explore the impact of lemma introduction on proof success rates, we extracted proof scripts that invoked previously defined lemmas during the proof process. This allows us to analyze whether existing proof synthesis tools struggle more with proofs that require the introduction of auxiliary lemmas.

2) Proof Pattern: Proof patterns in proof scripts refer to the recurring relationships between tactics within a proof sequence. These patterns capture the underlying logical reasoning connections between proof steps and reflect the typical strategies used in human-driven proofs. We used the PrefixSpan~\cite{han2001prefixspan,pei2004mining} sequential pattern mining algorithm to identify frequent patterns in proof strategies. These patterns were then classified by two authors of the paper based on their intended purpose in proofs. The inter-annotator agreement, measured by Cohen's Kappa coefficient, reached 0.90, indicating excellent consistency. Following discussions to resolve classification discrepancies, we ultimately categorized the proof patterns into six distinct types: Introduction, Simplification, Application, Rewriting, Automation, and Analysis. Subsequently, we conducted statistical analyses to determine the prevalence rates of these six proof patterns.

This experimental design aims to provide a comprehensive understanding of how structural characteristics of proof scripts, such as lemma usage and pattern, influence the success of proof synthesis tools. By examining these factors, we aim to identify patterns and insights that can guide the development of more effective proof automation strategies.

\subsubsection{RQ3: How does the search process in existing proof synthesis tools affect proof synthesis?\\}

\textbf{Motivation:} Existing proof synthesis tools predominantly adopt an approach that synthesizes individual proof steps sequentially and constructs complete proofs through iterative search. To achieve a more comprehensive evaluation of these tools, it is essential to conduct an in-depth analysis of their search mechanisms. Specifically, examining key elements of the search process, such as the number of proof attempts during the search and the types of proofs generated, can provide valuable insights. The number of attempts reflects the efficiency of the tool's search strategy, while the diversity and nature of the generated proofs reveal the tool's capability to explore the proof space effectively. Analyzing these factors can help identify bottlenecks in the search process and uncover patterns that contribute to successful proof synthesis. This research provides a foundation for understanding the limitations of current tools and guiding the development of more efficient and robust proof synthesis methodologies.

\textbf{Approach:} To study RQ3, we modified the code of four proof synthesis tools and reran their experiments to record detailed information about the proof search process. This included capturing the tactics used during the search, their corresponding tactic types, and other relevant metadata. Each proof's search process was ultimately represented by exploring a proof search tree, which provided a structured view of the tool's behavior during proof synthesis.

To analyze the relationship between the proof search process's characteristics and the tools' success rates, we calculated the average number of tactic attempts per proof step. This metric corresponds to the average number of child nodes for each parent node at every level of the proof search tree, offering insights into the search process's efficiency and branching behavior.

Additionally, we identified three distinct types of tactics encountered during the search process:

1) Error tactics: These tactics are directly rejected by the interactive theorem prover, resulting in an error message due to semantic invalidity.

2) Stagnation tactics: These tactics are accepted by the interactive theorem prover but generate proof subgoals that are redundant, duplicating previously proven goals and failing to advance the proof.

3) Progress tactics: These tactics successfully advance the proof by transforming the current goal into new, non-redundant subgoals.

To examine whether alignment with human expert proof patterns contributes to improved proof success, we conducted a detailed analysis of the proportion of human-like tactics within each of the three identified tactic types. We traverse the proof search tree and examine the relationships between parent and child nodes. For each node, we determine whether it conforms to a proof pattern. Using the previously defined three types of nodes, we then compute the proportion of pattern-conforming nodes within each type. This investigation aims to assess the potential of human-guided heuristics in distinguishing effective proof strategies from unproductive or erroneous ones, thereby offering insights into how human expert reasoning models might be leveraged to enhance the efficiency and reliability of automated proof search.

\section{Study Results and Analysis}
\label{sec:study_results}

This section presents the detailed results of our empirical study, providing an analysis of the limitations of existing proof synthesis tools from multiple perspectives.

\subsection{Results for RQ1}

\textbf{RQ1: Which characteristics of theorems make them harder to prove with existing proof synthesis tools?}

\begin{figure}[htbp]
\centering
\includegraphics[width=\linewidth]{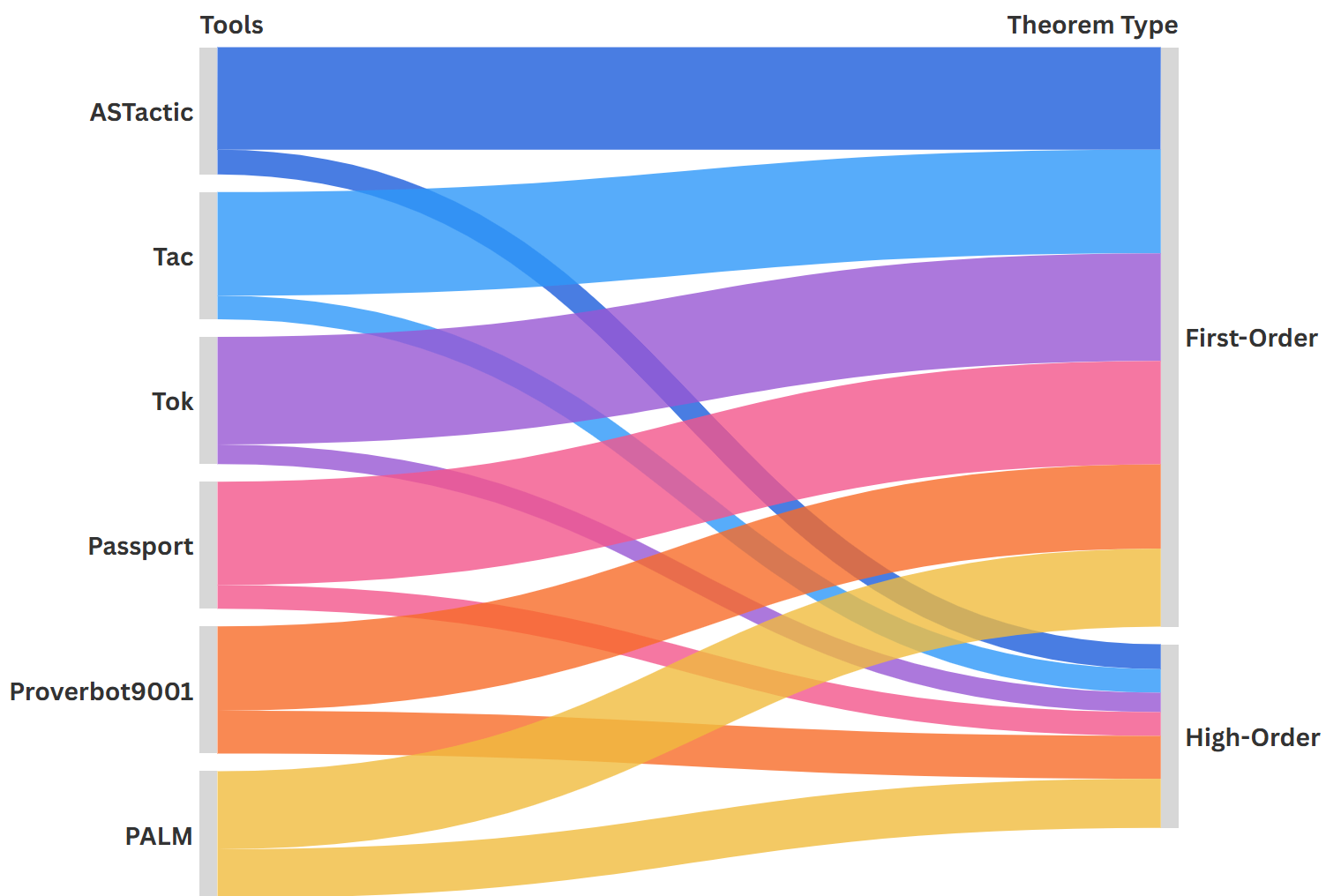}
\caption{The proportion of theorem types proven by proof synthesis tools}
\label{fig:type}
\end{figure}

To answer RQ1, we conduct a comprehensive analysis of performance bottlenecks in proof synthesis tools across two key dimensions: theorem types and complexity levels. Figure~\ref{fig:type} presents a Sankey diagram illustrating the distribution of theorem types proven by different proof synthesis tools. In the Sankey diagram, the left side represents different proof synthesis tools, while the right side represents the types of theorems. The flow from a synthesis tool to a theorem type indicates the proportion of theorems of each type among all theorems successfully proven by that tool. From the flow patterns in the diagram, we observe that the majority of the proof attempts by existing tools are concentrated in the first-order logic (FOL) block. This indicates that current tools are predominantly capable of proving first-order theorems, while higher-order theorems remain a significant challenge. 

Specifically, Tok exhibits the weakest performance on higher-order theorems, with only 15.3\% of its successfully proven theorems classified as higher-order. Although PALM achieves a relatively higher proportion of higher-order theorem proofs (38.6\%), the majority of the theorems it proves are still within the first-order domain. Further analysis of PALM's algorithm reveals that its success in proving a large number of theorems can be attributed to its repeated invocation of CoqHammer during the proof synthesis process. Since CoqHammer primarily leverages first-order automated reasoning techniques, this suggests that PALM’s performance is still largely dependent on first-order logic capabilities.

\begin{figure}[htbp]
\centering
\includegraphics[width=\linewidth]{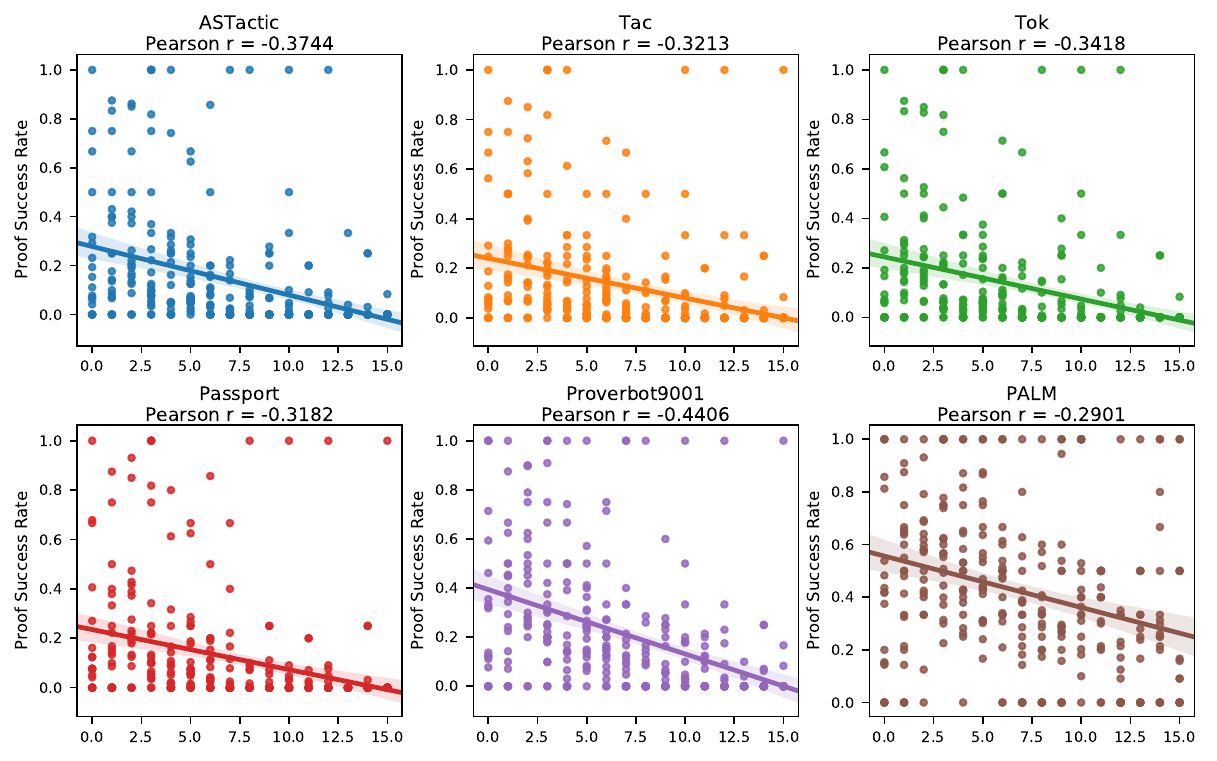}
\caption{Impact of Logical-Mathematical Symbol Count on Proof Success Rates of Proof Synthesis Tools}
\label{fig:complex}
\end{figure}

Figure~\ref{fig:complex} illustrates the relationship between the number of Logical-Mathematical Symbols in theorems and the proof success rate. The regression lines indicate that as the number of Logical-Mathematical Symbols increases, the proof success rates of all proof synthesis tools exhibit a consistent decline. This trend may be attributed to the increasing complexity and abstraction introduced by a higher density of symbols, which poses greater challenges for automated reasoning and proof construction. Among the tools evaluated, PALM demonstrates the highest stability, with a Pearson correlation coefficient closest to 0, suggesting minimal sensitivity to the increasing number of Logical-Mathematical Symbols. Conversely, Proverbot9001 exhibits the lowest stability, with a Pearson correlation coefficient approaching -1, indicating a strong negative correlation and significant vulnerability to the increasing complexity associated with a higher symbol count.

\rqbox{\textbf{Finding 1:} Proof synthesis tools exhibit significant performance bottlenecks as theorem complexity increases. Most tools, including PALM and Tok, perform well on first-order logic theorems but struggle with higher-order theorems and those containing a greater number of logical-mathematical symbols.}

\begin{figure}[htbp]
\centering
\includegraphics[width=\linewidth]{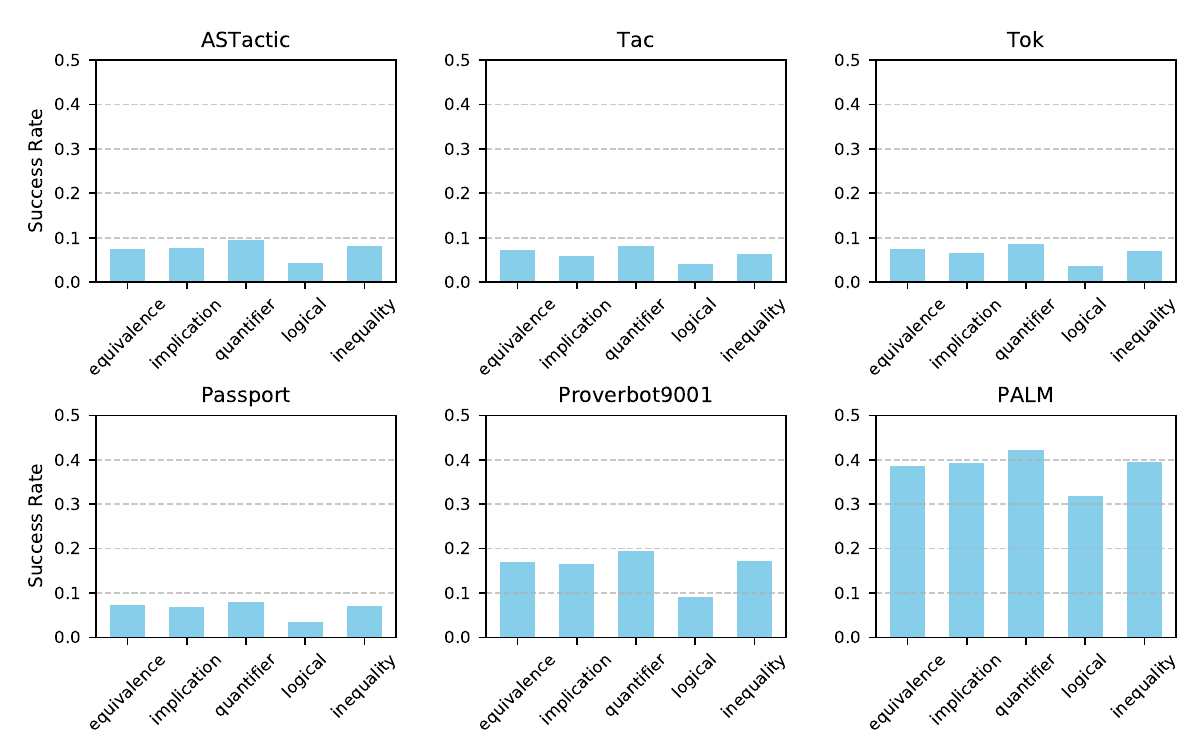}
\caption{Proof Success Rates of Theorems by Logical-Mathematical Symbol Type}
\label{fig:relation}
\end{figure}

Figure~\ref{fig:relation} illustrates the proof success rates of theorems involving different types of logical-mathematical symbols. Notably, theorems containing quantifiers (e.g., $\forall$, $\exists$) exhibit the highest proof success rate across all evaluated synthesis tools. In contrast, theorems involving implication operators ($\rightarrow$) and logical connectives ($\land$, $\lor$, $\lnot$) demonstrate relatively lower success rates. In particular, the presence of logical operators such as conjunction, disjunction, and negation appears to pose the greatest challenge, resulting in the lowest success rates among all categories. This suggests that the complexity introduced by logical connectives may hinder the performance of current automated proof synthesis techniques, possibly due to the increased reasoning steps and non-linear dependencies required to handle such structures.

\rqbox{\textbf{Finding 2:} Theorems with quantifiers (e.g., $\forall$, $\exists$) have the highest proof success rates across synthesis tools, while those with implication operators ($\rightarrow$) and logical operators ($\land$, $\lor$, $\lnot$) show lower success rates, with logical connectives posing the greatest challenge due to increased reasoning complexity.}

\subsection{Results for RQ2}

\textbf{RQ2: Which characteristics of proof scripts make them harder to prove with existing proof synthesis tools?}

\begin{figure}[htbp]
\centering
\includegraphics[width=\linewidth]{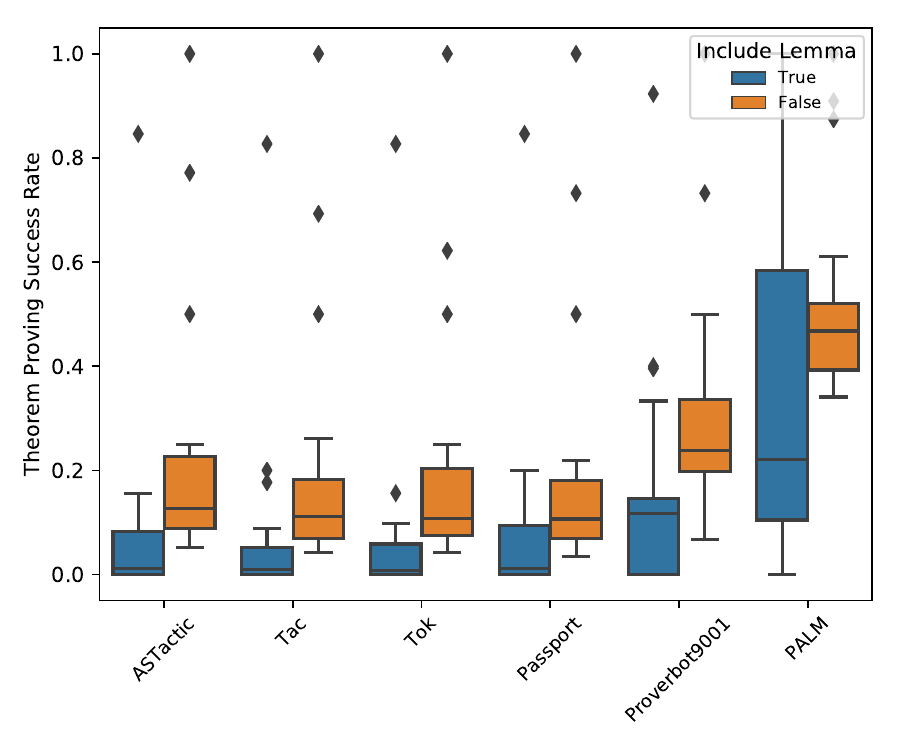}
\caption{Comparative Analysis of Proof Success Rates With and Without Auxiliary Lemma Introduction}
\label{fig:lemma}
\end{figure}

Figure~\ref{fig:lemma} presents a comparative analysis of proof success rates between scenarios involving the introduction of auxiliary lemmas and those without such additional components. The empirical data clearly demonstrates a consistent pattern across all proof synthesis tools: the success rates are significantly lower in cases requiring auxiliary lemmas compared to those that can be completed without lemma introduction. This systematic discrepancy suggests that the incorporation of auxiliary lemmas introduces substantial challenges for automated proof synthesis. Several underlying factors may contribute to this phenomenon: (1) the process of identifying and formulating appropriate auxiliary lemmas requires higher-level mathematical insight that current automated systems may lack; (2) the introduction of additional lemmas increases the complexity of the proof structure, potentially creating more opportunities for logical inconsistencies or implementation errors; (3) the need for auxiliary lemmas often indicates that the original problem is inherently more complex, which naturally leads to lower success rates. 

\rqbox{\textbf{Finding 3:} Automated proof synthesis tools exhibit significantly lower success rates on theorems requiring auxiliary lemmas, highlighting their limitations in handling structurally complex or insight-intensive proofs.}

\begin{figure}[htbp]
\centering
\includegraphics[width=\linewidth]{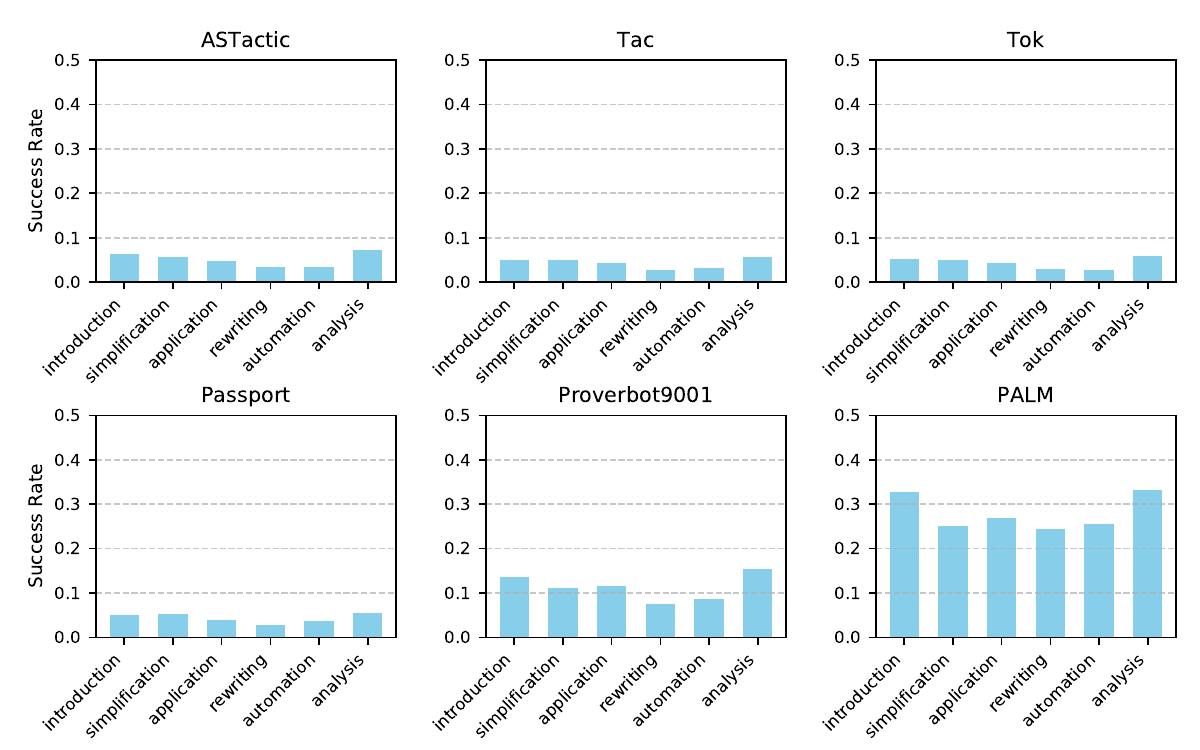}
\caption{Comparison of Proof Success Rates by Human Expert Proof Patterns}
\label{fig:pattern}
\end{figure}

Figure~\ref{fig:pattern} illustrates the proof success rates associated with different human expert proof patterns observed in the proof scripts. Among the various patterns, the "Introduction" and "Analysis" patterns exhibit the highest success rates, indicating that current proof synthesis tools are more effective when dealing with goal decomposition and case analysis tasks. In contrast, patterns such as "Rewriting" and "Automation" show comparatively lower success rates, suggesting that these techniques pose greater challenges for automated tools. This disparity may stem from the inherent complexity of applying rewriting rules and configuring automation tactics, which often require a deeper understanding of context-specific transformations and a careful selection of parameters or hint databases. 

\rqbox{\textbf{Finding 4:} Proof synthesis tools perform best on introduction and analysis patterns, but exhibit lower success rates on rewriting and automation, reflecting difficulties in managing context-sensitive transformations and tactic configuration.}

\subsection{Results for RQ3}

\textbf{RQ3: How does the search process in existing proof synthesis tools affect proof synthesis?}

\begin{figure*}[htbp]
\centering
\includegraphics[width=\textwidth]{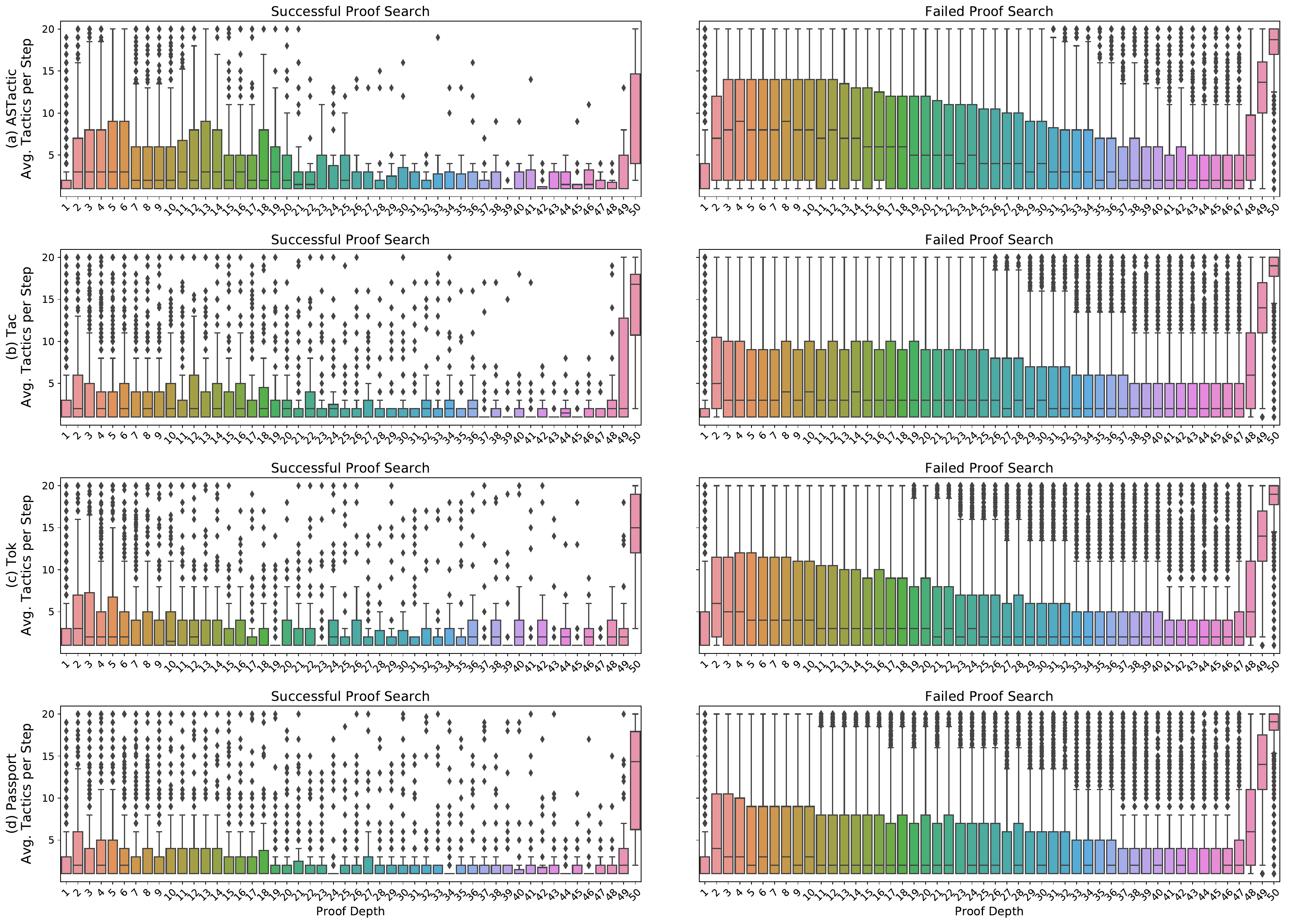}
\caption{Relationship Between Proof Search Depth and Number of Tactic Attempts for Successful and Failed Theorems}
\label{fig:node}
\end{figure*}

To address RQ3, we conducted an in-depth investigation into the characteristics of tactic usage during the proof search process. Specifically, we analyzed the relationship between search depth and the number of tactic attempts required to make progress in the proof. As shown in Figure~\ref{fig:node}, a significantly larger number of tactics are attempted during the initial stages of proof search, indicating that identifying a viable direction early in the proof is particularly challenging. As the proof progresses to greater depths, the number of required tactic attempts tends to decrease, suggesting that once a productive path is found, the search becomes more focused and efficient. 

Moreover, our analysis reveals a clear distinction between successful and failed proofs: theorems that ultimately fail to be proven consistently require a much higher number of tactic attempts throughout the search process. A closer inspection suggests that this is due to the model's tendency to assign high probabilities to tactics that either lead to errors or result in redundant proof states. As a consequence, the search must explore a larger number of low-ranked tactics to find one that successfully advances the proof. This behavior highlights a key limitation of current tactic prediction models: while they may be syntactically confident, their semantic alignment with proof progress is often weak, especially in complex or ambiguous proof states. These findings suggest that improving the accuracy and utility of tactic ranking, particularly in the early stages of proof search, is essential for enhancing proof synthesis performance.

\rqbox{\textbf{Finding 5:} Failed proof search involves significantly more tactic attempts, largely due to the model's overconfidence in tactics that lead to errors or redundant states, highlighting the need for more semantically informed tactic ranking.}

\begin{figure}[htbp]
\centering
\includegraphics[width=\linewidth]{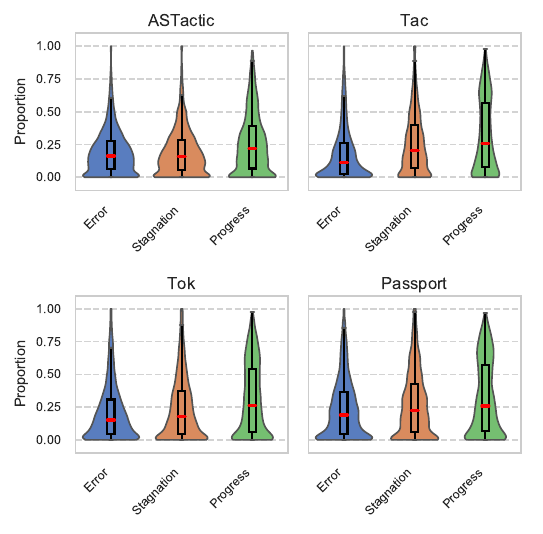}
\caption{Correlation of Human Expert Proof Pattern Alignment with Tactic Effectiveness in Proof Search}
\label{fig:pattern_combined}
\end{figure}

Figure~\ref{fig:pattern_combined} presents the proportion of tactics attempted during the proof search process that align with human expert proof patterns. Our analysis reveals that the proportion of tactics conforming to human expert proof patterns is significantly higher among those that successfully advance the proof compared to tactics that fail or result in redundant states. Specifically, tactics that advance the proof exhibit the highest alignment with human expert proof patterns, followed by those leading to redundant states, while tactics resulting in errors show the lowest alignment. Notably, both tactics that advance the proof and those leading to redundant states interact with the interactive theorem prover without triggering errors. These results indicate that alignment with human expert proof patterns strongly correlates with the syntactic and semantic validity of tactics in automated proof search.

\rqbox{\textbf{Finding 6:} Tactics that advance proofs in automated proof search show the highest alignment with human expert proof patterns, followed by those causing redundant states, while error-inducing tactics exhibit the lowest alignment, indicating a strong correlation between human expert proof pattern alignment and tactic validity.}


\section{Implications}
\label{sec:implications}

This section discusses the implications derived from our findings (Findings 1–6). We outline several insights that could inspire improvements for future proof synthesis tools.

\subsection{Design Implications for Automated Proof Synthesis Tools}

Automated proof synthesis tools must be capable of handling varying levels of theorem complexity and logical structure. Our study reveals several performance bottlenecks, highlighting the need for improvements in both tool architecture and tactic selection mechanisms. Based on these findings, we present a set of suggestions for enhancing future proof synthesis tools.

An effective direction for future proof synthesis tools is the joint prediction of tactics and theorem types, enabling more context-aware and accurate proof generation. We observe a notable decline in the performance of existing synthesis tools as theorem complexity increases (Finding 1). Although many SOTA tools (e.g., PALM and Tok) achieve high success rates on first-order logic theorems, their performance degrades significantly when faced with higher-order logic or theorems involving dense symbolic mathematical structures. This indicates that current tools struggle to perform effective tactic prediction for complex theorems. 

To overcome this limitation in Finding 1, one promising approach involves classifying proof tactics and labeling those that are typically used either individually or in combination in the proofs of complex theorems. For example, in Coq, complex proofs often involve tactics such as induction, inversion, apply, and rewrite, frequently combined with auxiliary lemmas and structured proof planning. Based on this observation, future synthesis tools could incorporate explicit complexity-awareness mechanisms. For instance, the tool could begin by classifying the input theorem according to its structural and symbolic complexity, and then dynamically adjust its tactic generation strategy accordingly. Furthermore, a multi-task learning-based tactic synthesizer could be trained to jointly predict both the next tactic and the complexity category of the current proof subgoal.

Future proof synthesis tools may require more fine-grained symbolic and semantic representation capabilities. Further analysis reveals that not all logical symbols contribute equally to proof difficulty. For instance, theorems involving quantifiers (e.g., $\forall$, $\exists$) tend to have higher proof success rates, whereas those involving implication ($\rightarrow$) and logical connectives such as conjunction ($\land$), disjunction ($\lor$), and negation ($\lnot$) show significantly reduced performance (Finding 2). In particular, the presence of logical connectives creates nonlinear dependency relationships and branched structures, which existing tools typically cannot handle efficiently. This limitation indicates a fundamental representational shortcoming in current logical encoding methods. 

Symbolic representation learning aims to encode the logical structure, symbolic relationships, and semantic roles of theorem components into vector or graph representations that can guide tactic prediction. When this representation is insufficient, tools may struggle to capture the complex interactions among symbols. This often leads to performance degradation, especially in the presence of implication and connectives. Existing methods may over-rely on syntactic patterns while overlooking deeper semantic roles. To overcome this challenge, future tools could incorporate semantically enriched embeddings. For example, the implication symbol ($\rightarrow$) could be labeled as a “hypothesis introduction” trigger, indicating its role in conditional reasoning. Such semantic labeling helps the model distinguish between syntactic forms that require different proof strategies. Additionally, the low success rate for implication-heavy theorems may stem from the model’s inability to learn how implication interacts with other symbols. As a simple example, the frequent co-occurrence of $\forall$ and $\rightarrow$ often corresponds to inductive or assumption-based proofs. Learning such interaction patterns can guide the model toward more suitable tactic choices in complex proof scenarios.

Current proof synthesis tools should incorporate lemma suggestion mechanisms as an essential component of their architecture. Our analysis reveals that many theorems require the introduction of intermediate lemmas to complete a successful proof. However, most existing systems lack the ability to identify or exploit such auxiliary lemmas, resulting in significantly lower success rates on theorems that inherently depend on intermediate reasoning steps (Finding 3). To overcome this limitation, future tools should include dedicated lemma discovery modules capable of automatically detecting when a particular proof state cannot be resolved by local tactics alone and instead requires the formulation of a supporting lemma. This involves not only recognizing the need for a lemma but also retrieving or generating a relevant lemma that can bridge the current proof gap. For example, in Coq, proving the associativity of list concatenation often requires the application of previously established lemmas about the identity element or the generalization of the induction hypothesis. These steps typically demand human expert insight or multi-step reasoning, which most current synthesis tools are not equipped to handle effectively. By integrating a lemma suggestion module, proof synthesis systems can more efficiently resolve intermediate proof states that would otherwise hinder progress. Such mechanisms are especially critical for theorems that require auxiliary insights, structural decomposition, or multi-hop reasoning. A well-designed lemma suggestion module has the potential to substantially improve performance in these more challenging cases, thereby narrowing the gap between automated and human-driven theorem proving.

Proof search strategies should be tailored according to the type of tactic pattern. Our analysis reveals that the tactic category used significantly impacts the automated proofs' success rate. In particular, introduction and analytic tactics yield higher success rates, while rewriting and automation-related tactics often underperform (Finding 4). This suggests that certain tactics, especially those requiring context-sensitive transformations or fine-grained configuration, pose greater challenges for existing automated proof synthesis tools. To enhance the robustness of proof automation across diverse tactic patterns, future tools may benefit from pattern-aware adjustments during proof search. This could involve dynamically selecting or prioritizing tactic sequences based on the recognized tactic type. For example, when the tools detect that a low-performing tactic category (e.g., rewriting) is likely needed, it could trigger selective tactic expansion, exploring multiple variants or alternatives more exhaustively. Additionally, constraint-based filtering mechanisms could be introduced to suppress low-confidence tactic applications, especially in cases where the model's prediction confidence does not meet a predefined threshold. Such approaches can reduce the search space and improve reliability by avoiding unproductive or misleading proof paths.

\subsection{Learning Implications for Human-Guided Proof Strategies}

While tool-level improvements are essential, a deeper integration of human-like reasoning patterns offers a complementary path forward. Our findings demonstrate that alignment with expert proof behavior significantly correlates with tactic validity and overall proof success, suggesting new directions for training and designing synthesis strategies.

Proof synthesis tools should incorporate human expert involvement or more advanced proof strategies to address model uncertainty. Our analysis reveals that proof failures are often associated with an increased number of tactic attempts, particularly when the model overconfidently selects tactics that lead to incorrect or redundant goals (Finding 5). This suggests that proof synthesis models exhibit high probabilistic uncertainty from the outset in cases where they eventually fail. To handle such uncertainty, greater human interaction should be incorporated when the model's predicted tactic distributions are highly uncertain. Alternatively, retrieval-augmented methods can be employed to improve the accuracy of tactic prediction by providing relevant contextual information from existing proofs. These approaches can help reduce failure rates, improve the robustness of automated proof synthesis, and allow the system to better leverage human intuition and domain expertise.

Human-written proof patterns can guide both training and proof search processes. We observe that tactics which successfully advance a proof are more likely to align with common human expert proof patterns, whereas tactics leading to redundant states or errors tend to exhibit lower alignment (Finding 6). This strong correlation suggests that consistency with human proof strategies can serve as a powerful inductive bias when training proof synthesis models. Future proof synthesis tools may leverage annotated human expert proof traces, employ imitation learning from expert tactic sequences, or incorporate preference modeling based on human interaction logs. Additionally, prioritizing tactics that align with human expert proof patterns during proof search may further facilitate the synthesis of complete proof scripts. These approaches not only enhance the quality of tactic selection but also improve the interpretability of the proof process and its adaptability to user feedback.

\section{Automated Proof Synthesis via PGTS}
\label{sec:method}

Existing proof synthesis tools typically generate tactics for individual proof steps without considering the logical reasoning relationships between them. However, human expert reasoning patterns that reflect the dependencies and logical progression between proof steps are often crucial to constructing coherent proofs. According to RQ3, proofs that align with human-written patterns tend to be more successful in the search process. This finding inspires us to consider whether guiding proof search with such patterns can improve success rates. To explore this, we investigate pattern-guided tactical search strategies to enhance existing depth-first search methods. Specifically, we first extract common proof patterns from a large-scale corpus of human-written proofs and formalize these patterns into reusable templates. In this way, we integrate the mined human expert proof patterns into the proof search process, providing heuristic guidance to steer the search more effectively.

\subsection{Mining Proof Pattern}

To obtain human-written proof patterns, we follow the methodology outlined in RQ2 for extracting proof patterns. First, we convert 57,719 human-written proof scripts from the CoqGym training set into sequential data. In this step, each tactic in a proof script is treated as an individual item in a sequence, and the tactics are recorded in the order they are executed. Furthermore, we normalize the tactic sequences by unifying tactic parameters (i.e., simplifying parameterized tactics into their parameter-free forms) and handling compound tactics (i.e., decomposing tactic chains separated by semicolons, such as intros; simpl; auto, into individual items). Through this preprocessing, we obtain standardized sequences of proof tactics.

Next, we apply the PrefixSpan sequential pattern mining algorithm to extract human-written proof patterns. We follow a common practice in setting the pattern mining threshold. Specifically, we extract patterns that appear with a frequency greater than 1\% in the dataset~\cite{pereira2015mining}. Since proof scripts typically consist of a series of ordered tactics and exhibit temporal or event sequences characteristics, the PrefixSpan algorithm is well-suited for this task. PrefixSpan focuses on mining frequent subsequences by recursively projecting the sequence database to discover patterns, thereby avoiding generating an excessive number of candidate sequences. Since the PrefixSpan algorithm mines all frequent subsequences, it may include non-contiguous patterns with large spans. Therefore, to ensure that the mined patterns are strictly contiguous, we apply a filtering step to retain only strictly contiguous patterns.

\subsection{Pattern-Guided Tactic Search}

The proof synthesis tool is designed to generate candidate tactics for advancing each proof step. It employs a step-by-step search strategy to construct complete proof sequences. The proof search can be modeled as a tree search, where each node represents a candidate tactic generated by the automated synthesis tool, and a path from the root to a leaf corresponds to a potential complete proof sequence. This search process is formally described in Algorithm~\ref{alg:pgts}, which integrates neural predictions with mined human expert proof patterns to guide and prioritize tactic application during proof construction.

\begin{algorithm*}[htbp]
\caption{PGTS: Pattern-Guided Tactic Search}
\label{alg:pgts}
\KwIn {Initial goal \( G \), Pattern frequency table \( P \), Neural network predictor \( NN \), Top-\( k = 20 \)}
\KwOut {Proof sequence or "No proof found"}

\Fn{ProofSearch(\( G, P, NN \))}{
    \( Proof \gets [] \) \tcp*{Proof sequence}
    \( Stack \gets \text{push\_sorted}(\MyCall{PGTS}{G, \text{null}, P, NN})\) \tcp*{Push init top-k tactics, top-1 at stack top}
    \( Seen \gets \{ G \} \) \tcp*{Seen goals to detect repetition}
    \While{\( Stack \neq [[]] \)}{
        \If{\( Stack[-1] = [] \)}{
            \( Stack.\text{pop}() \), \( Proof.\text{pop}() \), \( \MyCall{Undo}{} \) \tcp*{Backtrack}
            \Continue
        }
        
        \( Tactic \gets Stack[-1].\text{pop}() \) \tcp*{get tactic to apply current goal}
        \( (Result, G) \gets \MyCall{ApplyTactic}{Tactic, G} \) \tcp*{prove result and next proof goal}
        
        \If{\( Result = \text{"SUCCESS"} \)}{
            \( Proof \gets Proof \cup \{ Tactic \} \) \tcp*{Add tactic to proof sequences}
            \Return \( Proof \) \tcp*{Proof found}
        }
        \ElseIf{\( Result = \text{"MAX\_NUM\_TACTICS\_REACHED"} \) \textbf{or} \( Result = \text{"MAX\_TIME\_REACHED"} \)}{
            \Return "No proof found" \tcp*{Fail if limits reached}
        }
        \ElseIf{\( Result = \text{"ERROR"} \) \textbf{or} \( G \in Seen \)}{
            \Continue \tcp*{Skip if error or goal repeated}
        }
        \ElseIf{\( Result = \text{"PROVING"} \)}{
            \( Proof \gets Proof \cup \{ Tactic \} \) \tcp*{Add tactic}
            \( Seen.\text{add}(G) \) \tcp*{Mark goal as seen}
            \( Stack.\text{push}(\MyCall{PGTS}{G, Tactic, P, NN}) \) \tcp*{Push new tactics for next proof goal}
        }
    }
    \Return "No proof found" \tcp*{Proof not found}
}

\Fn{PGTS(\( G, ParentTactic, P, NN \))}{
    \( T \gets NN(G) \) \tcp*{Top-\( k \) tactics predicted by neural network}
    \If{\( ParentTactic \) is \( \text{null} \)}{
        \Return \( T \) \tcp*{Return NN predictions if no prior tactic}
    }
    \( T_{pattern} \gets \{ t \in T \mid (ParentTactic, t) \in P \} \) \tcp*{Pattern-matching tactics}
    \( T_{rest} \gets T \setminus T_{pattern} \) \tcp*{Remaining tactics}
    \Return \( \MyCall{SortByPattern}{T_{pattern}, P} \cup \MyCall{SortByNN}{T_{rest}} \) 
    \tcp*{Sort pattern-matching tactics by frequency, others by NN probability}
}
\end{algorithm*}

Specifically, The proof process begins with an initial goal derived from the theorem to be proven. At this point, a neural network is used to generate the top-k candidate tactics that most likely apply to the current goal (Line 3). The value of k for the top-k candidate tactics is set to 20, following the default configuration used in prior work. Since the initial goal has no associated parent tactic, it cannot be matched against any binary tactic transition patterns mined from human proofs (Lines 32-34). Consequently, the candidate tactics are attempted strictly in descending order of the probabilities output by the neural network.

The system attempts to apply each tactic in the top-k list using the interactive theorem prover (Lines 10-11). If the application of a tactic results in an error or if the new proof goal it generates has already been seen (Lines 16-18), the system skips to the next tactic in the list. When a tactic successfully advances the proof by producing a new, unseen proof goal, it is added to the proof sequence and recorded as the parent tactic for future pattern matching (Lines 22–26). It is important to note that if none of the top-k candidate tactics generated for a given goal successfully advance the proof, the system performs backtracking (Lines 6-9). In this case, it returns to the preceding proof goal and resumes the search by attempting the remaining, previously untried tactics associated with the parent node. This mechanism ensures that the search does not terminate prematurely and can explore alternative proof paths when an initially chosen tactic sequence fails.

The new proof goal then undergoes the same process: the neural network predicts a new set of top-k tactics. This time, however, our Pattern-Guided Tactic Search (PGTS) algorithm re-ranks the neural network outputs based on frequent tactic transition patterns previously mined from human-written proofs (Lines 35-37). PGTS reorders the candidate tactics in two stages. First, it identifies tactics that match a known binary pattern with the current parent tactic. These pattern-matching tactics are sorted in descending order of their frequency in the pattern database. Second, any tactics that do not match a known pattern are sorted by the original neural network confidence scores. The re-ranked list combines both subsets, prioritizing pattern-aligned tactics while still considering the model's statistical predictions.

The proof search continues by sequentially trying the reordered tactics with the interactive theorem prover. If a tactic successfully leads to proof completion, meaning no subgoals remain, the proof is considered successful, and the system returns the accumulated proof sequence (Lines 12–15). The process is terminated and reported as unsuccessful if the proof search exceeds a predefined maximum number of tactic attempts or a time threshold (Lines 16-17). We follow the settings used in prior work by imposing a maximum time limit of 600 seconds and restricting the total number of tactic applications to 300. These constraints ensure a fair comparison with existing approaches and prevent the search process from running indefinitely.
\section{Evaluation}
\label{sec:evaluation}

Our evaluation answers three research questions:
\begin{itemize}
    \item \textbf{RQ4 (Effectiveness):} Does PGTS enhance the effectiveness of theorem proving?
    \item \textbf{RQ5 (Capability):} Does PGTS enhance the ability of proof synthesis tools to generate complex theorems?
    \item \textbf{RQ6 (Quality):} Does PGTS improve the quality of proof generation?
\end{itemize}

\subsection{Evaluation Setup}

\subsubsection{Benchmark}

\textbf{Dataset.} We use CoqGym, a widely adopted benchmark for evaluating theorem-proving synthesis tools. The CoqGym benchmark consists of 124 open-source Coq projects. Its training set includes 57,719 theorems from 97 projects, while the test set contains 13,137 theorems from 27 projects. We extract common human-written proof patterns from the proof sequences in the training set and evaluate their effectiveness on the test set. 

\textbf{Base model.} Our goal is to develop a general proof search optimization method. We selected four proof synthesis tools, ASTactic, Tac, Tok, and Passport, as baseline models. All of these utilize a DFS proof search strategy. Additionally, to better evaluate the effectiveness of our proposed approach, we implemented Restricted Pattern-Guided Tactic Search (RPGTS), a variant of PGTS that prioritizes only parameterized tactics that match predefined patterns.

\subsubsection{Evaluation Metrics}

For RQ4, we use Total Added Value and Unique Added Value to evaluate the extent to which PGTS enhances the proof capabilities of existing proof synthesis tools. The detailed definitions of these two metrics are provided below. 

\textbf{Total Added Value:} This metric shows how much better tool A performs compared to tool B by calculating the extra theorems A proves (total theorems by A minus total theorems by B) divided by the total theorems B proves. It indicates the overall percentage improvement of A over B in proving theorems.

\textbf{Unique Added Value:} This metric measures the specific advantage of tool A over tool B by determining the number of theorems only A proves (not proven by B) divided by the total theorems B proves. It reflects the percentage of new theorems A adds relative to B’s performance.

For RQ5, we evaluate the improvement in the ability of proof synthesis tools to handle complex theorems by measuring the proportion of newly proven theorems that involve higher-order logic. For RQ6, we assess the enhancement in the quality of generated proofs by calculating the average number of proof steps in the synthesized scripts. Since variations in the number and complexity of theorems may introduce bias into the average proof length, we introduce an additional metric: the average reduction in proof steps compared to human-written proofs. This metric reflects the conciseness of the synthesized scripts and provides a more detailed perspective on proof quality.

\subsection{RQ4: Effectiveness}

\begin{figure}[htbp]
\centering
\includegraphics[width=0.5\textwidth]{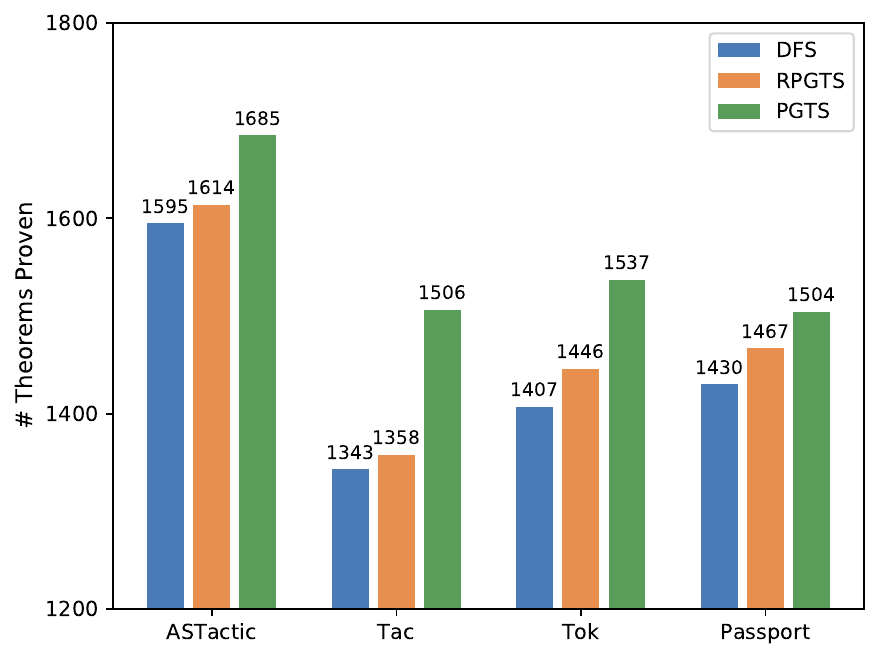}
\caption{Number of Theorems Proved by Baseline Tools Using DFS, PGTS, and RPGTS}
\label{fig:total}
\end{figure}

Figure~\ref{fig:total} illustrates the number of theorems proved by all studied baselines using the original depth-first search (DFS), our proposed Pattern-Guided Tactic Search (PGTS), and a variant, RPGTS. The results show that incorporating human-written proof patterns into the search process as a heuristic significantly enhances the theorem-proving capabilities of automated proof synthesis tools. Notably, guiding all tactics with patterns, as in PGTS, outperforms guiding only parameterized tactics, as in RPGTS. Specifically, PGTS improves the ASTactic tool by 5.64\% (90 additional theorems), Tac by 12.14\% (163 additional theorems), Tok by 9.24\% (130 additional theorems), and Passport by 5.17\% (74 additional theorems), achieving an average performance improvement of 8.05\% across proof synthesis tools.

Following the evaluation methodology of previous theorem proving tools, we also investigated whether incorporating pattern-guided search enhances the ability of proof synthesis tools to prove new theorems. Table~\ref{tab:unique} presents the unique added value of our proposed pattern-guided approach (PGTS) and a variant method (RPGTS) compared to the original DFS proof search. Consistent with the total added value results, employing pattern-guided proof search significantly enhances the ability of theorem-proving synthesis tools to prove new theorems. Guiding all tactics with patterns, as in PGTS, substantially outperforms guiding only parameterized tactics, as in RPGTS. Specifically, PGTS proves 240 theorems that the original ASTactic fails to prove, yielding a unique added value of 15.05\%. For Tac, PGTS proves 327 additional new theorems, achieving a unique added value of 24.35\%. For Tok, it proves 329 new theorems, with a unique added value of 23.38\%, and for Passport, it proves 246 new theorems, resulting in a unique added value of 17.20\%. Overall, PGTS proves an average of 286 theorems that prior tools fail to address, with an average unique added value of 20.00\%. These results underscore PGTS's superior ability to uncover previously unprovable theorems by effectively leveraging comprehensive pattern guidance, outperforming both DFS and the more limited RPGTS approach.

\begin{table}[htbp]
\setlength{\tabcolsep}{2pt} 
\caption{Unique Added Value of PGTS and RPGTS Compared to Original DFS Proof Search Across Theorem-Proving Tools}
\label{tab:unique}
\begin{tabular}{l c c c}
\toprule
Base Model & DFS & \begin{tabular}{@{}c@{}}RPGTS \\ Unique Added Value\end{tabular} & \begin{tabular}{@{}c@{}}PGTS \\ Unique Added Value\end{tabular} \\
\midrule
ASTactic   & 1595 & 50 (3.13\%)  & 240 (15.05\%) \\
Tac        & 1343 & 68 (5.06\%)  & 327 (24.35\%) \\
Tok        & 1407 & 94 (6.68\%)  & 329 (23.38\%) \\
Passport   & 1430 & 82 (5.73\%)  & 246 (17.20\%) \\
\midrule
Average    & -    & 74 (5.15\%)  & 286 (20.00\%) \\
\bottomrule
\end{tabular}
\end{table}

\rqbox{\textbf{Answer to RQ4:} PGTS improves the performance of proof synthesis tools by increasing the number of successfully proven theorems by an average of 8.05\%. On average, it proves 20\% more theorems that were previously unprovable by the baseline tools.}

\subsection{RQ5: Capability}

Table~\ref{tab:new_complex} presents the proportion of higher-order logic in the newly proven theorems by PGTS, indicating that PGTS proves a greater number of new theorems involving higher-order logic. The results show that across all four proof synthesis tools, PGTS consistently proves a higher percentage of new theorems involving higher-order logic compared to the original DFS search. Specifically, for ASTactic, the proportion of higher-order logic new theorems increases from 19.44\% under DFS to 29.17\% with PGTS, marking a 50.05\% relative increase. Tac exhibits an even more significant improvement, with the proportion rising from 18.76\% to 32.72\%, reflecting a 74.41\% increase. Similarly, Tok sees a substantial increase from 16.84\% under DFS to 38.91\% with PGTS, achieving a remarkable 131.00\% relative gain. Passport also demonstrates a strong improvement, with the proportion increasing from 20.42\% to 37.80\%, representing an 85.11\% increase. On average, PGTS leads to an 85.14\% increase in the proportion of proven theorems that involve higher-order logic across all tools. These findings indicate that PGTS is particularly effective in addressing complex theorems that require higher-order reasoning, significantly enhancing the capability of automated proof synthesis tools.

\begin{table}[htbp]
\centering
\caption{Proportion of Higher-Order Logic in New Proven Theorems by PGTS}
\label{tab:new_complex} 
\setlength\tabcolsep{9pt}
\begin{tabular}{lccc} 
\toprule 
Base Model & DFS & new Theorems & Increase \\ 
\midrule 
ASTactic & 19.44\% & 29.17\% & 50.05\% \\
Tac      & 18.76\% & 32.72\% & 74.41\% \\
Tok      & 16.84\% & 38.91\% & 131.00\% \\
Passport & 20.42\% & 37.80\% & 85.11\% \\
\midrule 
Average   & -     & -     & 85.14\% \\ 
\bottomrule 
\end{tabular}
\end{table}

\rqbox{\textbf{Answer for RQ5:} PGTS significantly enhances the ability of various proof synthesis tools to prove new theorems involving higher-order logic, with an average increase of 85.14\%.}

\subsection{RQ6: Quality}

\begin{figure}[htbp]
\centering
\includegraphics[width=0.5\textwidth]{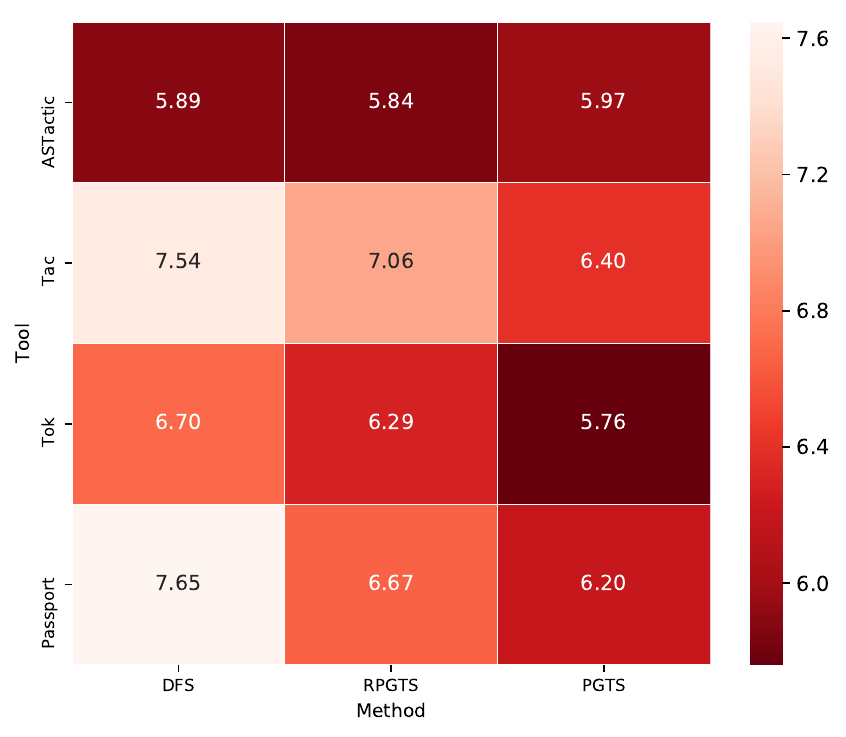}
\caption{The Average Proof Sequence Length of Four Proof Synthesis Tools Using Different Proof Search Methods}
\label{fig:predictionstep}
\end{figure}

Figure~\ref{fig:predictionstep} presents a heatmap illustrating the average proof length generated by different proof search strategies across four foundational proof synthesis tools. In the heatmap, darker colors indicate shorter proof sequences, signifying higher-quality proofs. The experimental results demonstrate that our proposed PGTS tool significantly improves proof synthesis by reducing proof lengths in most cases. Specifically, PGTS achieves an average proof length reduction of 15.12\% in Tac, 14.03\% in Tok, and 18.95\% in Passport compared to the baseline DFS strategy. For instance, in Passport, PGTS generates proofs with an average length of 6.20 tactics, outperforming DFS (7.65 tactics) and RPGTS (6.67 tactics), thereby demonstrating a notable improvement. However, on ASTactic, PGTS produces slightly longer proof sequences than the initial DFS search strategy employed by the tool. Nonetheless, its variant, RPGTS, still generates shorter proofs than DFS. This suggests that pattern-guided proof search remains effective, as it leverages structural regularities in proof synthesis to guide the search toward more efficient proof sequences.

Since different proof synthesis tools address distinct theorems, the average proof length may be influenced by the complexity of the theorems involved. To address this, we also evaluated the quality of proof sequences generated by these tools by measuring the difference between the proof sequence length produced by the tools and the length of manually written proof sequences. Figure 3 presents the reduction in proof step length achieved by different search methods across four proof synthesis tools, with darker heatmap colors indicating a smaller proof length compared to manually written proofs. 
Our analysis reveals that the average step length reduction for current proof synthesis tools is consistently negative. This suggests that the proof sequences generated by these tools are generally longer than their manually written counterparts. Furthermore, our proposed PGTS search method demonstrates a notable improvement over the original DFS search method, achieving a greater reduction in proof sequence length. For instance, PGTS reduces proof lengths by an average of 20.82\% (0.97 tactics steps) more than DFS across the four tools, highlighting its superior performance in producing more concise proofs.

\begin{figure}[htbp]
\centering
\includegraphics[width=0.5\textwidth]{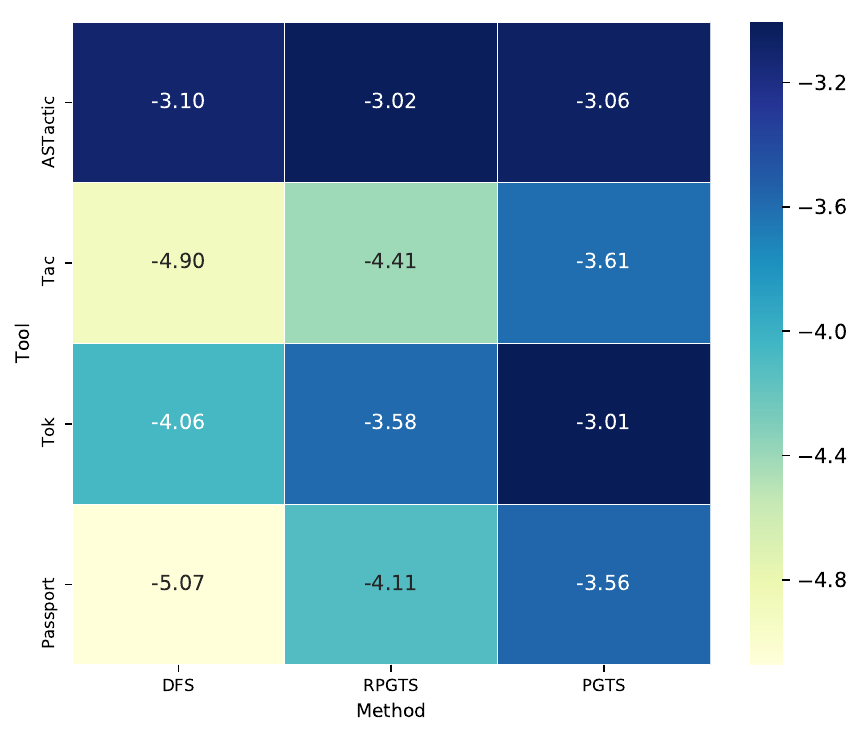}
\caption{The Average Reduction in Proof Steps by Four Proof Synthesis Tools Using Different Proof Search Methods Compared to Human-Written Proofs}
\label{fig:proofreduction}
\end{figure}

\rqbox{\textbf{Answer for RQ6:} PGTS significantly reduces proof lengths compared to DFS, achieving an average reduction of 20.82\% across proof synthesis tools.}

\section{Discussion}
\label{sec:discussion}

This section discusses several potential limitations of our approach PGTS, as well as threats to its validity.

\subsection{Limitations}

\begin{figure}[htbp]
\centering
\includegraphics[width=0.5\textwidth]{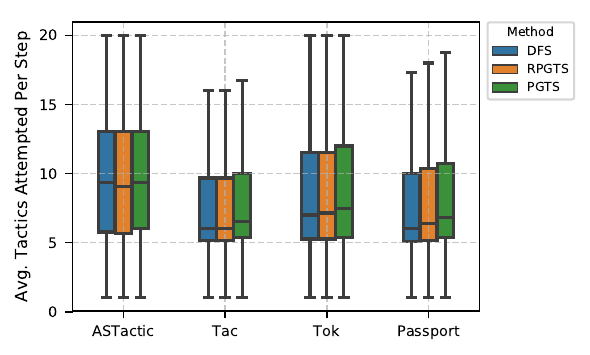}
\caption{Average Number of Tactic Attempts per Proof Step under Different Search Strategies}
\label{fig:attempt}
\end{figure}

While our proposed PGTS significantly improves the capability of existing proof synthesis tools in handling complex theorems and generating more concise proof scripts, we observed a notable increase in the number of tactic attempts during the proof search process. Figure~\ref{fig:attempt} illustrates the number of tactics attempted at each proof step when using the original DFS-based search compared to our pattern-guided approaches, RPGTS and PGTS. Interestingly, PGTS tends to attempt slightly more tactics per step than DFS.

To understand this phenomenon, we conducted a deeper analysis. The primary cause lies in the pattern-matching mechanism: tactics that align with human-like proof patterns are prioritized during the search. However, our current design considers only the category of a tactic, without validating the correctness of its parameters before attempting it. As a result, many tactics that superficially match the desired pattern but have incorrect parameters are explored first, leading to an increase in tactic attempts.

Despite this limitation, guiding the proof search with human-like patterns still significantly improves the effectiveness of proof synthesis. Prioritizing tactics that align with common proof strategies helps the search process remain more focused and increases the likelihood of finding successful proofs, even if some additional tactic attempts are required. However, it is important to note that directly filtering tactics based on parameter validity remains highly challenging. Verifying whether a tactic's parameters are correct typically requires interaction with the underlying ITPs, as static analysis alone is insufficient to determine parameter correctness in most cases. Such interactions introduce considerable computational overhead and would fundamentally change the lightweight nature of the current search framework. Consequently, it is difficult to filter out invalid tactics prior to their execution. PGTS, therefore, strikes a practical balance by guiding search using tactic categories informed by human-like patterns without engaging in expensive parameter validation. Future improvements may explore lightweight heuristic approximations or predictive models to estimate parameter correctness without fully relying on ITPs interactions, though this remains an open challenge.

\subsection{Threats to Validity}

\textbf{Internal Validity.} Threat to internal validity lies in the potential bias during tactic pattern identification. PGTS relies on tactic patterns derived from empirical data, and any inaccuracies in the extraction process may impact its generalizability. To mitigate this risk, we applied a sequential pattern mining algorithm to extract tactic patterns from manually written proof scripts. To ensure consistency, the extracted patterns are restricted to strictly contiguous sequences of tactics.


\textbf{External Validity.} Threats to external validity arise from the generalizability of both our empirical study and the experimental results of the PGTS approach. Although we evaluated six state-of-the-art proof synthesis tools, all of them are specifically designed for the Coq interactive theorem prover. This Coq-specific focus may limit the applicability of our findings to other proof assistants. However, our empirical results demonstrate a consistent pattern of behavior across all six tools, suggesting that the observed phenomena may reflect broader characteristics of current proof synthesis methods. Furthermore, the fact that our PGTS approach enhances the proof capabilities of all evaluated tools provides additional evidence supporting its generalizability beyond any single tool.

\section{Related Works}
\label{sec:rel}

\subsection{Neural Theorem Proving}
Neural theorem proving integrates neural language models with formal proof assistants, leveraging large-scale corpora of existing proofs to train the system to generate proof tactics automatically. Yang et al.~\cite{yang2019learning} employed a deep learning-based model to generate abstract syntax trees for tactic sequences, demonstrating the potential of neural networks in generating tactics for interactive theorem provers. First et al.~\cite{first2020tactok} further integrated partially written proof scripts with proof state modeling, enhancing the success rate of automatically generated tactics for interactive theorem proving. Sanchez-Stern et al.~\cite{sanchez2023passport} enhanced the capabilities of existing proof synthesis tools by incorporating identifier encoding. Furthermore, First et al.~\cite{first2022diversity} demonstrated that the proof success rate can be significantly improved by combining multiple complementary machine learning-based tactic generation models. 

With the advancement of large model technology, some research~\cite{kasibatla2024cobblestone,lu2024proof,jiang2022thor} has explored the integration of large models with symbolic-based first-order theorem provers, such as Hammer~\cite{paulsson2012three,kaliszyk2014learning,czajka2018hammer}, for proof generation. First et al.~\cite{first2023baldur} achieved promising results by fine-tuning large models for the generation and repair of proofs. Yang et al.~\cite{yang2023leandojo} proposed a retrieval-augmented method for selecting premises from extensive mathematical libraries to facilitate proof synthesis. 

\subsection{Empirical Study on Code-related Tasks}
The advancement of deep learning technologies has led to the emergence of numerous techniques tailored to code-related tasks. These include empirical evaluation of vulnerability detection tools, code summarization, and other software engineering applications. Existing empirical studies can be categorized into two types: one explores the impact of dataset quality on model performance, while the other focuses on the capability boundaries of models for specific tasks. Liu et al.~\cite{liu2018neural} found that removing noise from the commit message dataset significantly degraded the model's performance. Wu et al.~\cite{wu2021data} found that correcting mislabeled security bug reports improved the performance of the classification model. Chakraborty et al.~\cite{chakraborty2021deep} demonstrated that state-of-the-art deep learning-based techniques experienced a performance drop of more than 50\% in real-world vulnerability prediction scenarios. Tian et al.~\cite{tian2022makes} indicated that the dataset for the commit message generation task has issues with message quality and defined what constitutes good commit quality. Unlike the aforementioned tasks, the proof script synthesis dataset is of high quality, as it is derived from real runtime proof information and evaluated on theorems from real-world projects.

Peng et al.\cite{peng2022revisiting} conducted a systematic study on the performance of methods for the API recommendation task. Dong et al.\cite{dong2023revisiting} examined the frequent patterns in commit messages generated by models. Steenhoek et al.~\cite{steenhoek2023empirical} investigated the types of program vulnerabilities that deep learning models may find challenging to handle. Zhang et al.\cite{zhang2024automatic} studied the types of commit messages generated and demonstrated that current methods can only generate commit message subjects rather than message bodies. Ni et al.\cite{ni2024learning} investigated seven research questions on vulnerability detection models across multiple dimensions, including model capability, interpretability, stability, usability, and cost-effectiveness. Similar to these works, this paper explores proof synthesis tools' capability boundaries. We analyze the limitations of these tools from three perspectives: the theorem itself, proof scripts, and proof search. Furthermore, based on our observations, we propose a pattern-guided tactic search method to optimize existing tools, significantly improving proof performance.

\section{Conclusion}
\label{sec:con}
In this paper, we explored the limitations of current proof synthesis tools based on deep learning within ITPs environments. We observed that tactics aligned with human expert proof patterns are more likely to be those that successfully advance the proof during the search process. This finding motivated the development of our PGTS approach, which integrates pattern-awareness into the proof search process of existing proof synthesis tools. PGTS heuristically prioritizes tactics that match human expert proof patterns, resulting in more efficient and focused proof search. Evaluation results demonstrate that PGTS significantly enhances existing proof synthesis tools, increasing the average theorem-proving rate by 8.05\%. Moreover, PGTS proves to be especially effective in handling complex theorems and generating more concise proof sequences, suggesting its potential as a general improvement framework for a wide range of ITPs.

\section*{Acknowledgment}
This work was supported by the Major Program of the National
Natural Science Foundation of China (Grant Nos. 62192733
and 62192730), the General Program of National Natural Science Foundation of China No.62372219.

\bibliographystyle{unsrtnat}
\bibliography{reference}

@inproceedings{yang2019learning,
  title={Learning to prove theorems via interacting with proof assistants},
  author={Yang, Kaiyu and Deng, Jia},
  booktitle={International Conference on Machine Learning},
  pages={6984--6994},
  year={2019},
  organization={PMLR}
}

@article{first2020tactok,
  title={TacTok: Semantics-aware proof synthesis},
  author={First, Emily and Brun, Yuriy and Guha, Arjun},
  journal={Proceedings of the ACM on Programming Languages},
  volume={4},
  number={OOPSLA},
  pages={1--31},
  year={2020},
  publisher={ACM New York, NY, USA}
}

@article{sanchez2023passport,
  title={Passport: Improving automated formal verification using identifiers},
  author={Sanchez-Stern, Alex and First, Emily and Zhou, Timothy and Kaufman, Zhanna and Brun, Yuriy and Ringer, Talia},
  journal={ACM Transactions on Programming Languages and Systems},
  volume={45},
  number={2},
  pages={1--30},
  year={2023},
  publisher={ACM New York, NY}
}

@inproceedings{sanchez2020generating,
  title={Generating correctness proofs with neural networks},
  author={Sanchez-Stern, Alex and Alhessi, Yousef and Saul, Lawrence and Lerner, Sorin},
  booktitle={Proceedings of the 4th ACM SIGPLAN International Workshop on Machine Learning and Programming Languages},
  pages={1--10},
  year={2020}
}

@inproceedings{lu2024proof,
  title={Proof automation with large language models},
  author={Lu, Minghai and Delaware, Benjamin and Zhang, Tianyi},
  booktitle={Proceedings of the 39th IEEE/ACM International Conference on Automated Software Engineering},
  pages={1509--1520},
  year={2024}
}

@inproceedings{first2022diversity,
  title={Diversity-driven automated formal verification},
  author={First, Emily and Brun, Yuriy},
  booktitle={Proceedings of the 44th International Conference on Software Engineering},
  pages={749--761},
  year={2022}
}

@article{kasibatla2024cobblestone,
  title={Cobblestone: Iterative Automation for Formal Verification},
  author={Kasibatla, Saketh Ram and Agarwal, Arpan and Brun, Yuriy and Lerner, Sorin and Ringer, Talia and First, Emily},
  journal={arXiv preprint arXiv:2410.19940},
  year={2024}
}

@inproceedings{first2023baldur,
  title={Baldur: Whole-proof generation and repair with large language models},
  author={First, Emily and Rabe, Markus N and Ringer, Talia and Brun, Yuriy},
  booktitle={Proceedings of the 31st ACM Joint European Software Engineering Conference and Symposium on the Foundations of Software Engineering},
  pages={1229--1241},
  year={2023}
}

@article{jiang2022thor,
  title={Thor: Wielding hammers to integrate language models and automated theorem provers},
  author={Jiang, Albert Qiaochu and Li, Wenda and Tworkowski, Szymon and Czechowski, Konrad and Odrzyg{\'o}{\'z}d{\'z}, Tomasz and Mi{\l}o{\'s}, Piotr and Wu, Yuhuai and Jamnik, Mateja},
  journal={Advances in Neural Information Processing Systems},
  volume={35},
  pages={8360--8373},
  year={2022}
}

@article{yang2023leandojo,
  title={Leandojo: Theorem proving with retrieval-augmented language models},
  author={Yang, Kaiyu and Swope, Aidan and Gu, Alex and Chalamala, Rahul and Song, Peiyang and Yu, Shixing and Godil, Saad and Prenger, Ryan J and Anandkumar, Animashree},
  journal={Advances in Neural Information Processing Systems},
  volume={36},
  pages={21573--21612},
  year={2023}
}

@inproceedings{paulsson2012three,
  title={Three years of experience with Sledgehammer, a practical link between automatic and interactive theorem provers},
  author={Paulsson, Lawrence C and Blanchette, Jasmin C},
  booktitle={Proceedings of the 8th International Workshop on the Implementation of Logics (IWIL-2010), Yogyakarta, Indonesia. EPiC},
  volume={2},
  year={2012}
}

@article{kaliszyk2014learning,
  title={Learning-assisted automated reasoning with Flyspeck},
  author={Kaliszyk, Cezary and Urban, Josef},
  journal={Journal of Automated Reasoning},
  volume={53},
  pages={173--213},
  year={2014},
  publisher={Springer}
}

@article{czajka2018hammer,
  title={Hammer for Coq: Automation for dependent type theory},
  author={Czajka, {\L}ukasz and Kaliszyk, Cezary},
  journal={Journal of automated reasoning},
  volume={61},
  pages={423--453},
  year={2018},
  publisher={Springer}
}

@inproceedings{pereira2015mining,
  title={Mining comparative sentences from social media text},
  author={Pereira, Fab{\'\i}ola SF},
  booktitle={Workshop on interactions between data mining and natural language processing (DMNLP) co-located with European conference on machine learning and principles and practice of knowledge discovery in databases (ECML/PKDD)},
  pages={41--48},
  year={2015}
}

@inproceedings{tian2022makes,
  title={What makes a good commit message?},
  author={Tian, Yingchen and Zhang, Yuxia and Stol, Klaas-Jan and Jiang, Lin and Liu, Hui},
  booktitle={Proceedings of the 44th International Conference on Software Engineering},
  pages={2389--2401},
  year={2022}
}

@inproceedings{dong2023revisiting,
  title={Revisiting learning-based commit message generation},
  author={Dong, Jinhao and Lou, Yiling and Hao, Dan and Tan, Lin},
  booktitle={2023 IEEE/ACM 45th International Conference on Software Engineering (ICSE)},
  pages={794--805},
  year={2023},
  organization={IEEE}
}

@article{peng2022revisiting,
  title={Revisiting, benchmarking and exploring API recommendation: How far are we?},
  author={Peng, Yun and Li, Shuqing and Gu, Wenwei and Li, Yichen and Wang, Wenxuan and Gao, Cuiyun and Lyu, Michael R},
  journal={IEEE Transactions on Software Engineering},
  volume={49},
  number={4},
  pages={1876--1897},
  year={2022},
  publisher={IEEE}
}

@article{zhang2024automatic,
  title={Automatic commit message generation: A critical review and directions for future work},
  author={Zhang, Yuxia and Qiu, Zhiqing and Stol, Klaas-Jan and Zhu, Wenhui and Zhu, Jiaxin and Tian, Yingchen and Liu, Hui},
  journal={IEEE Transactions on Software Engineering},
  volume={50},
  number={4},
  pages={816--835},
  year={2024},
  publisher={IEEE}
}

@inproceedings{steenhoek2023empirical,
  title={An empirical study of deep learning models for vulnerability detection},
  author={Steenhoek, Benjamin and Rahman, Md Mahbubur and Jiles, Richard and Le, Wei},
  booktitle={2023 IEEE/ACM 45th International Conference on Software Engineering (ICSE)},
  pages={2237--2248},
  year={2023},
  organization={IEEE}
}

@article{ni2024learning,
  title={Learning-based models for vulnerability detection: An extensive study},
  author={Ni, Chao and Shen, Liyu and Xu, Xiaodan and Yin, Xin and Wang, Shaohua},
  journal={arXiv preprint arXiv:2408.07526},
  year={2024}
}

@article{chakraborty2021deep,
  title={Deep learning based vulnerability detection: Are we there yet?},
  author={Chakraborty, Saikat and Krishna, Rahul and Ding, Yangruibo and Ray, Baishakhi},
  journal={IEEE Transactions on Software Engineering},
  volume={48},
  number={9},
  pages={3280--3296},
  year={2021},
  publisher={IEEE}
}

@article{wu2021data,
  title={Data quality matters: A case study on data label correctness for security bug report prediction},
  author={Wu, Xiaoxue and Zheng, Wei and Xia, Xin and Lo, David},
  journal={IEEE Transactions on Software Engineering},
  volume={48},
  number={7},
  pages={2541--2556},
  year={2021},
  publisher={IEEE}
}

@inproceedings{liu2018neural,
  title={Neural-machine-translation-based commit message generation: how far are we?},
  author={Liu, Zhongxin and Xia, Xin and Hassan, Ahmed E and Lo, David and Xing, Zhenchang and Wang, Xinyu},
  booktitle={Proceedings of the 33rd ACM/IEEE international conference on automated software engineering},
  pages={373--384},
  year={2018}
}

@inproceedings{klein2009sel4,
  title={seL4: Formal verification of an OS kernel},
  author={Klein, Gerwin and Elphinstone, Kevin and Heiser, Gernot and Andronick, June and Cock, David and Derrin, Philip and Elkaduwe, Dhammika and Engelhardt, Kai and Kolanski, Rafal and Norrish, Michael and others},
  booktitle={Proceedings of the ACM SIGOPS 22nd symposium on Operating systems principles},
  pages={207--220},
  year={2009}
}

@article{leroy2009formal,
  title={Formal verification of a realistic compiler},
  author={Leroy, Xavier},
  journal={Communications of the ACM},
  volume={52},
  number={7},
  pages={107--115},
  year={2009},
  publisher={ACM New York, NY, USA}
}

@inproceedings{staples2014productivity,
  title={Productivity for proof engineering},
  author={Staples, Mark and Jeffery, Ross and Andronick, June and Murray, Toby and Klein, Gerwin and Kolanski, Rafal},
  booktitle={Proceedings of the 8th ACM/IEEE International Symposium on Empirical Software Engineering and Measurement},
  pages={1--4},
  year={2014}
}

@inproceedings{leroy2016compcert,
  title={CompCert-a formally verified optimizing compiler},
  author={Leroy, Xavier and Blazy, Sandrine and K{\"a}stner, Daniel and Schommer, Bernhard and Pister, Markus and Ferdinand, Christian},
  booktitle={ERTS 2016: Embedded Real Time Software and Systems, 8th European Congress},
  year={2016}
}

@article{barthe2019formal,
  title={Formal verification of a constant-time preserving C compiler},
  author={Barthe, Gilles and Blazy, Sandrine and Gr{\'e}goire, Benjamin and Hutin, R{\'e}mi and Laporte, Vincent and Pichardie, David and Trieu, Alix},
  journal={Proceedings of the ACM on Programming Languages},
  volume={4},
  number={POPL},
  pages={1--30},
  year={2019},
  publisher={ACM New York, NY, USA}
}

@article{zhangsurvey,
  title={A Survey of Formal Verification Approaches for Practical Systems},
  author={Zhang, Qiao and Zhuo, Danyang and Wilcox, James}
}

@inproceedings{ferreira2021reliability,
  title={Reliability in software-intensive systems: challenges, solutions, and future perspectives},
  author={Ferreira, Francisco Henrique and Nakagawa, Elisa Yumi and dos Santos, Rodrigo Pereira},
  booktitle={2021 47th Euromicro Conference on Software Engineering and Advanced Applications (SEAA)},
  pages={54--61},
  year={2021},
  organization={IEEE}
}

@article{ferreira2023towards,
  title={Towards an understanding of reliability of software-intensive systems-of-systems},
  author={Ferreira, Francisco Henrique Cerdeira and Nakagawa, Elisa Yumi and dos Santos, Rodrigo Pereira},
  journal={Information and Software Technology},
  volume={158},
  pages={107186},
  year={2023},
  publisher={Elsevier}
}

@book{kapur2011software,
  title={Software reliability assessment with OR applications},
  author={Kapur, PK and Pham, Hoang and Gupta, Anshu and Jha, PC and others},
  volume={364},
  year={2011},
  publisher={Springer}
}

@inproceedings{lyu2007software,
  title={Software reliability engineering: A roadmap},
  author={Lyu, Michael R},
  booktitle={Future of Software Engineering (FOSE'07)},
  pages={153--170},
  year={2007},
  organization={IEEE}
}

@inproceedings{han2001prefixspan,
  title={Prefixspan: Mining sequential patterns efficiently by prefix-projected pattern growth},
  author={Han, Jiawei and Pei, Jian and Mortazavi-Asl, Behzad and Pinto, Helen and Chen, Qiming and Dayal, Umeshwar and Hsu, Meichun},
  booktitle={proceedings of the 17th international conference on data engineering},
  pages={215--224},
  year={2001},
  organization={IEEE Piscataway, NJ, USA}
}

@article{pei2004mining,
  title={Mining sequential patterns by pattern-growth: The prefixspan approach},
  author={Pei, Jian and Han, Jiawei and Mortazavi-Asl, Behzad and Wang, Jianyong and Pinto, Helen and Chen, Qiming and Dayal, Umeshwar and Hsu, Mei-Chun},
  journal={IEEE Transactions on knowledge and data engineering},
  volume={16},
  number={11},
  pages={1424--1440},
  year={2004},
  publisher={IEEE}
}

@misc{pgts,
  author = {Zhang, Manqing and Zhou, Lingru and Xiao, Bingxu and Dong, Yunwei and Liu, Yepang},
  title = {Replication package for “Understanding and Improving Automated Proof Synthesis for Interactive Theorem Provers”},
  year = {2025},
  howpublished = {\url{https://github.com/zmqgeek/PGTS}}
}


\end{document}